\documentclass[aps,twocolumn,amsmath,showpacs,pra]{revtex4}
  \usepackage[dvips]{color}
  \usepackage{psfig,newlfont}
  \usepackage{graphicx}
  \usepackage{amssymb}
  \usepackage{amsmath}
  \usepackage[latin1]{inputenc}
  \usepackage{float}

\begin{document}
\title{Quantum NOT Operation and Integrability in Two-Level Systems}
\author{R. M. Angelo}
\affiliation{Universidade Federal do Paraná, Departamento de Física, Caixa Postal 19044, Curitiba 81531 990, PR, Brazil.}
\email{renato@fisica.ufpr.br}
\author{W. F. Wreszinski}
\affiliation{Universidade de São Paulo, Instituto de Física,\\ 
Caixa Postal 66318, São Paulo 05315 970, SP, Brazil.}

\begin{abstract}
We demonstrate the surprising integrability of the classical Hamiltonian associated to a spin $1/2$ system under periodic external fields. The one-qubit rotations generated by the dynamical evolution is, on the one hand, close to that of the rotating wave approximation (RWA), on the other hand to two different ``average'' systems, according to whether a certain parameter is small or large. Of particular independent interest is the fact that both the RWA and the averaging theorem are seen to hold well beyond their expected region of validity. Finally we determine conditions for the realization of the quantum NOT operation by means of classical stroboscopic maps.

\pacs{03.67.Lx, 42.50.Ct, 05.45-a}

\end{abstract}

\maketitle

\section{Introduction}
Advanced NMR techniques are able to manipulate qubits in order to implement several quantum logic operations. In particular, by means of radio-frequency magnetic fields, nuclear spin $1/2$ can be controlled to the realization of the quantum NOT-gate, one of the fundamental logic block in quantum computation \cite{vander04}. On the other hand, in the field of quantum optics, the advent of strong laser pulses brings up the possibility of investigating the fundamentals of the radiation-matter interaction. In this scenario, models like a two-level quantum system interacting with an external classical field have achieved a paradigmatic place, since it has been succefully applied by the theorists to make predictions in NMR \cite{vander04}, quantum computation \cite{divicenzo95} and quantum optics \cite{hioe84}.

From a mathematical point of view its importance emerges mainly from the low dimensionality of its Hilbert space, which allows analytical analysis of the dynamics in many cases of interest. For periodic external fields, e.g., such systems have been solved analytically only in the rotating wave approximation (RWA), which restricts the problem to both resonance and weak coupling regimes. Far off this situation, sophisticated perturbative methods have been applied \cite{barata00,wresz98,frasca03} and exact solutions have been found only for specific periodic fields \cite{bagrov01}. An alternative approach for studying exactly the two-level problem is provided by the picture of the classical gyromagnet \cite{feynman}. The dynamics of the matrix elements of a pure density operator is mapped on a classical Hamiltonian and methods of dynamical systems may be employed to study the dynamics of the system. Alternatively, it is possible to define a classical dynamics over the Bloch sphere \cite{allen}.

In this work we focus on these classical frameworks to study relevant questions as the integrability of the underlying classical dynamics, the validity of the RWA (as well as its derivation from the averaging method) and regimes for the realization of a unitary NOT operation. The paper is organized as follows. In section \ref{sec:classical} we define the two classical frameworks on which the analysis will take place. In section \ref{sec:integrability} we demonstrate the integrability of the classical dynamics associated to unitary two-level systems under arbitrary external fields. In sections \ref{sec:R} and \ref{sec:NR} we present several results concerning the phase space aspect of two important fields commonly used in experiments. In section \ref{NOTgate} we use the results offered by the stroboscopic maps to make predictions about the quantum NOT operation. Section \ref{sec:conclusion} is reserved for some concluding remarks.

\section{The Classical Framework}\label{sec:classical}
The dynamics of a spin 1/2 system interacting with an external time-dependent magnetic field $\mathbf{B}(t)$ in the dipole approximation is given by the Hamiltonian
\begin{eqnarray}
H(t)=-\frac{1}{2}\mathbf{B}(t)\cdot\mathbf{\Sigma}, \label{HQ}
\end{eqnarray}
in which $\mathbf{\Sigma}=(\sigma_1,\sigma_2,\sigma_3)$ is the vector composed by the Pauli matrices. For simplicity we have adopted $\hbar=1$, so that magnetic field is given in units of frequency. The dynamics of the density operator $\rho(t)=|\psi(t)\rangle\langle\psi(t)|$ is governed by the von Neumann equation, $\imath \dot{\rho}=[H(t),\rho]$.

As mentioned, according to Feynman {\it et al} \cite{feynman}, it is possible to formulate the dynamics in terms of a classical Hamiltonian systems. 
Consider the following parametrization for the quantum dynamics in terms of a classical vector $(S_0,\mathbf{S})$:
\begin{eqnarray}
\rho(t)=\frac{1}{2}\left(S_0\,\mathbf{1}+\mathbf{S}\cdot\mathbf{\Sigma} \right)=\frac{1}{2}
\left(\begin{array}{cc} S_0+S_3 & S_1-\imath S_2 \\ S_1+\imath S_2 & S_0-S_3   \end{array} \right), \label{rhoS}
\end{eqnarray}
in which $\mathbf{1}$ stands for the unity matrix.
The trace $(\textrm{Tr}\rho=1)$ and purity $(\rho^2=\rho)$ conditions impose that $S_0=1$ and $\mathbf{S}^2=S_1^2+S_2^2+S_3^2=1$, respectively. The quantum equation of motion for $\rho$ is then transformed in a system of differential equations described by
\begin{eqnarray}
\frac{d\mathbf{S}}{dt}=\mathbf{S}\times\mathbf{B}, \label{edS}
\end{eqnarray}
which may be obtained from the classical Hamiltonian
\begin{eqnarray}
\cal{H}(t)=-\mathbf{B}(t)\cdot \mathbf{S}. \label{HC}
\end{eqnarray}
Equations \eqref{edS} and \eqref{HC} define the classical geometric picture associated to quantum two-level systems: the unit vector $\mathbf{S}$ precesses around the vector $\mathbf{B}$ just like a {\it classical gyromagnet} precesses in a magnetic field.

Following reference \cite{bagrov01} we consider the unit sphere $\cal{S}^2$ with the usual angular coordinates $0\leqslant\theta\leqslant\pi$, $0\leqslant \varphi \leqslant 2\pi$, and let $\mathbf{S}=(\sin\theta\cos\varphi,\sin\theta\sin\varphi,\cos\theta)$. Introducing $p=\varphi$ and $q=-\cos\theta$ as canonically conjugate variables, we may write
\begin{eqnarray}
\mathbf{S}=(\sqrt{1-q^2}\cos{p},\sqrt{1-q^2}\sin{p},-q), \label{Sqp}
\end{eqnarray}
with the usual Poisson brackets $\{S_1,S_2\}=S_3$ (plus ciclic permutations).
By \eqref{HC} and \eqref{Sqp} we finally write
\begin{eqnarray}
\cal{H}=-\left[B_1(t)\cos{p}+B_2(t)\sin{p}\right]\sqrt{1-q^2}-B_3(t) q,
\label{Hqp}
\end{eqnarray}
with equations of motion given by
\begin{subequations}
\begin{eqnarray}
\dot{q}&=&\left[B_1 \sin{p}-B_2\cos{p} \right]\sqrt{1-q²}, \\ 
\dot{p}&=&-\left[B_1\cos{p}+B_2 \sin{p}\right] \frac{q}{\sqrt{1-q²}}-B_3,
\end{eqnarray}
\end{subequations}
in which the amplitudes $B_i$ are functions of time. 

In our analysis it will be useful to express the quantum states in terms of the classical canonical pair $(q,p)$. Consider a general quantum state of a single qubit as $|\psi\rangle=a|+\rangle+b e^{-\imath \phi}|-\rangle$ being $a$ and $b$ positive real numbers satisfying $a²+b²=1$ and $\phi$ a the relative phase. Composing the corresponding density operator, $\rho=|\psi\rangle\langle\psi|$, and comparing with \eqref{rhoS} and \eqref{Sqp}, one may write
\begin{eqnarray}
|\psi\rangle=\sqrt{\frac{1-q}{2}}|+\rangle+\sqrt{\frac{1+q}{2}}e^{\imath p}|-\rangle. \label{psiqp}
\end{eqnarray}
Thus, correspondence between classical phase space and quantum states becomes immediate. For instance, for $q=\mp 1$, we have $S_3=\pm 1$ and $|\psi\rangle=|\pm \rangle$, with obvious interpretation. Notice that the relative phase of the quantum state has the status of a conjugate momentum in classical phase space. It indeed emphasizes its dynamical relevance.

We will also consider a perpendicular state $|\psi_{\bot}\rangle$, defined by $\langle\psi_{\bot}|\psi\rangle=0$. According to \eqref{psiqp} it reads
\begin{eqnarray}
|\psi_{\bot}\rangle=\sqrt{\frac{1+q}{2}}|+\rangle+\sqrt{\frac{1-q}{2}}e^{\imath (p+\pi)}|-\rangle. \label{psibot}
\end{eqnarray}
Thus, the transformation
\begin{eqnarray}
|\psi\rangle \rightarrow |\psi_{\bot}\rangle,
\end{eqnarray}
implies in the following classical counterpart
\begin{eqnarray}
(p,q) \rightarrow (p+\pi,-q).
\end{eqnarray}
By \eqref{Sqp} we see that such a condition of orthogonality implies that $\mathbf{S}\to-\mathbf{S}$. These considerations will be essential for our analysis of the quantum NOT operation in two-level systems.
\section{Integrability} \label{sec:integrability}
The classical Hamiltonian \eqref{Hqp} may be written in a two-dimensional autonomous form by means of Howland's method \cite{reed75}. In this case, since the new Hamiltonian is composed by an integrable part added to a non-integrable perturbation, chaotic behavior is expected to be present \cite{bagrov01}. However, as we show now, the unitarity of the underlying quantum dynamics guarantees that this is not the case. Consider the following distance in the Bloch sphere:
\begin{eqnarray}
D(t)\equiv ||\,S_1(t)-S_2(t)\, ||,\label{DS}
\end{eqnarray}
where different indices refer to different initial conditions. By \eqref{rhoS} it is easy to show that $D$ may also be written as
\begin{eqnarray}
D(t)=\sqrt{2\,\textrm{Tr}\left[\rho_1(t)-\rho_2(t) \right]²},\label{Drho}
\end{eqnarray}
where different indices refer to different initial states $\rho_{1,2}(0)$. Since the dynamics is unitary, we may write
\begin{eqnarray}
\rho_{1,2}(t)=U(t)\,\rho_{1,2}(0)\,U^{\dag}(t),\label{rho12}
\end{eqnarray}
in which $U$ is the propagator satisfying $\imath\,\dot{U}=H(t)\,U(t)$. Equation \eqref{DS} allows us to define a Lyapunov exponent as
\begin{eqnarray}
\lambda=\lim\limits_{D(0)\to 0}\,\lim\limits_{t\to\infty}\frac{1}{t}\ln\left[\frac{D(t)}{D(0)} \right], \label{lambda}
\end{eqnarray}
which measures the mean exponential departure between two arbitrarily close initial conditions on the Bloch sphere. By \eqref{rho12} and \eqref{Drho} it is immediate that $D(t)=D(0)$ and thus, by \eqref{lambda}, $\lambda=0$. When a full set (i.e, equal to the number of degrees of freedom) of constants of the motion in involution does not exist, so that we have a system which is not integrable, sensivity to initial conditions is certain to exist at least in some region pf the phase space, i.e., there is inevitably chaotic behavior. Thus we may regard the proof that $\lambda=0$ as a proof of integrability, although it may be difficult to find explicitly an additional constant of the motion for the general autonomous system equivalent to \eqref{HC} (see, e.g., \cite{bagrov01}, for the description of these autonomous systems). It is worth emphasizing that our proof of integrability is valid for an arbitrary time-dependent field, including the quasiperiodic case. This contradicts the basic assertion of \cite{pomeau} and confirms the result of \cite{badii}. More importantly, we show the real reason for the integrability observed in \cite{badii} using double Poincaré sections. Notice also that a field in \eqref{HC} depending of $N$ incommensurate frequencies, would yield, by Howland's method \cite{reed75}, a $N+1$ degrees of freedom autonomous system. Our demonstration asserts that such a system is integrable for all $N$, a surprising result.

We close this section with some remarks which may help to connect the present model with the fundamental model of interaction between a (two-level) atom and the quantized electromagnetic field in the dipole approximation \cite{nus73}, which we shall take as one-mode for simplicity:
\begin{eqnarray}
H=\omega\,a^{\dag}a\otimes\mathbf{1}+\frac{\omega_0}{2}\,\mathbf{1}\otimes\sigma_z+g\,\left(a+a^{\dag}\right)\otimes\sigma_x,
\label{HFQ}
\end{eqnarray}
where $a$ and $a^{\dag}$ are the annihilation and creation operators, respectively, and $\mathbf{1}$ is the identity matrix (properly defined in each subspace). Although ``quantum integrability'' is not a well-defined concept in general, two natural definitions, with direct analogue in classical mechanics are possible: 1) \eqref{HFQ} is integrable if there exists a unitary transformation $U_1$ such that $U_1HU_1^{-1}=H(N,\sigma_z)$, where $N=a^{\dag}a$ and $\sigma_z$ are two ``quantum actions'' (commuting operators, whose spectrum and eigenvectors are known explicitly); 2) \eqref{HFQ} is integrable if there exists a unitary transformation $U_2$ such that the spin and bosonic degrees of freedom decouple. It has been proved in \cite{amniat} that \eqref{HFQ} is integrable according to 1) for both large and small coupling $g$. In reference \cite{lo} it has been remarked that
\begin{eqnarray}
\tilde{H}&=&U_2^{\dag}\,H\,U_2\nonumber \\
&=&\frac{\omega_0}{2}\sigma_z\,\cos{\left(\pi a^{\dag}a \right)}+\omega\, a^{\dag}a+g\left(a^{\dag}+a \right),
\label{Htilde}
\end{eqnarray}
where
\begin{eqnarray}
U_2=\exp{\left\{-\imath\pi\left(\sigma_x-\mathbf{1} \right)a^{\dag}a/2 \right\}}.
\label{U2}
\end{eqnarray}
Thus, by \eqref{Htilde}, in each sector corresponding to the eigenvalues $\pm 1$ of $\sigma_z$ the system is equivalent to a one-dimensional model (equivalent to one degree of freedom in classical mechanics), and thus the spin and bosonic degrees of freedom are decoupled. Notice, however, that the resulting model is highly ``nonsoluble''! This fact reflects upon the semiclassical limit of \eqref{HFQ}, in which the ``mean photon number'' (properly defined) tends to infinity, keeping the photon density fixed \cite{guerin}: it yields our model \eqref{HQ}, with $\mathbf{B}=\mathbf{B}_{NR}$ given by \eqref{BNR}, whose dynamics is known, since the seminal papers of Bloch and Siegert \cite{bloch} and Autler and Townes \cite{autler}, to be highly nontrivial. For the system \eqref{HFQ} there exist two commuting operators: $H$, given by \eqref{HFQ}, and the parity operator
\begin{eqnarray}
\Pi=\exp{\left\{\imath\pi\left[a^{\dag}a+\frac{1}{2}\left(\sigma_z+\mathbf{1} \right) \right] \right\}}.
\label{parity}
\end{eqnarray}
The fact that $\Pi$ corresponds to a discrete symmetry without classical analogue might have suggested that the semiclassical limit \eqref{HQ}, with \eqref{BNR}, is not integrable, which we have seen not to be the case.

Above, we have considered one atom only. A different limit which may be performed on \eqref{HFQ} is the ``many-atom limit'' (properly defined \cite{hepp}). The method of \cite{hepp} justifies rigorously the results of \cite{graham,milonni,aguiar}, whereby the resulting coupled classical equations may be shown to display chaotic behaviour.

\section{The Rotating Field} \label{sec:R}
In this section we will focus our analysis in a special case of periodic external field: the radio-frequency field \cite{vander04}. This field, which we will call {\it rotating} (R) in virtue of a rotation symmetry around the direction $3$, is defined as
\begin{eqnarray}
\mathbf{B}_{R}(t)=-2 \Big(B_0 \cos{(\omega t+\phi)}, B_0 \sin{(\omega t+\phi)},B_3\Big),\label{BR}
\end{eqnarray}
being $\phi$ the phase of the field and $B_0$ and $B_3$ constant amplitudes. By \eqref{Hqp} and \eqref{BR}, we write the corresponding Hamiltonian as
\begin{eqnarray}
\cal{H}(q,p,t)=2 B_0\sqrt{1-q²} \cos{(p+\phi-\omega t)}-2 B_3 q.
\label{HR}
\end{eqnarray}
The integrability of this model may be explicitly verified by means of a canonical transformation that produces an autonomous one-dimensional Hamiltonian. Consider the following generating function and respective transformation equations \cite{goldstein}:
\begin{subequations}
\begin{eqnarray}
F_3(p,Q,t)&=&-(p+\phi)\,Q+\theta(t)\,Q, \\
q&=&-\frac{\partial F_3}{\partial p}=Q,\label{qQ} \\
P&=&-\frac{\partial F_3}{\partial Q}=p+\phi-\theta,\label{pP} \\
\cal{K}&=&\cal{H}+\frac{\partial F_3}{\partial t}= \cal{H}+\dot{\theta} 
Q,\label{KH}
\end{eqnarray} \label{F3}
\end{subequations}

\noindent
being $\theta(t)=\omega t$ in this case.
It is just the classical counterpart of the usual unitary quantum rotation given by $\exp{(-\imath \omega t \sigma_3)}$. Equations above yield
\begin{eqnarray}
\cal{K}(Q,P)=2 B_0\sqrt{1-Q²}\cos{P}-2\,\Omega\, Q,
\label{K}
\end{eqnarray}
with $\Omega = B_3-\frac{\omega}{2}$. Now we have an integrable Hamiltonian, since $\cal{K}$ is a constant of motion in a one-dimensional system. Therefore, the system R defined by \eqref{HR} is manifestly integrable. Nonetheless, because the phase space $(p,q)$ contains separatrices (as we will see latter) finding analytical solutions for the equations of motion may not be a simple task. The simplest (and more reliable) way of getting the solutions is solving the problem in the original quantum system and then implementing the corresponding transformations. We present the analytical solutions in the appendix \ref{app:analytical}. In the next sections we will deal with the explicit solutions. For now it is sufficient for our purposes to understand the general characteristics of this integrable system.
 
Before constructing the stroboscopic maps, we point out one further interesting information. By \eqref{F3}:
\begin{eqnarray}
\cal{H}=\cal{K}(q_0,p_0)-\omega q,\label{EqS}
\end{eqnarray}
or
\begin{eqnarray}
\frac{d\cal{H}}{dq}=\omega, \label{dHdq}
\end{eqnarray}
which indicates that the time-dependent energy, $\cal{H}$, keeps a linear relation with $q$ for all times.

\subsection{Stroboscopic Map}\label{sec:map}
In the general case of a one-dimensional Hamiltonian system driven by an external periodic field, for the which analytical analysis is not possible, stroboscopic maps are useful tools to attest the integrability. In our case, although we have already proved the integrability, it will be useful constructing the stroboscopic maps in order to obtain some sights about the whole aspect of the integrable tori. As we will see, it will be essential in our analysis of the NOT operation.

The stroboscopic map is obtained by marking the position and its conjugate momentum at each multiple of the Hamiltonian period, $T$, which is defined by the relation $\cal{H}(q,p,t+T)=\cal{H}(q,p,t)$. 
A rapid inspection of the Hamiltonian (\ref{HR}) yields $T=\frac{2\pi}{\omega}$. We then define the stroboscopic instants as $t_k=k T$, for $k\in\mathbb{N}$.

\begin{figure}[ht]
\includegraphics[scale=0.45,angle=0]{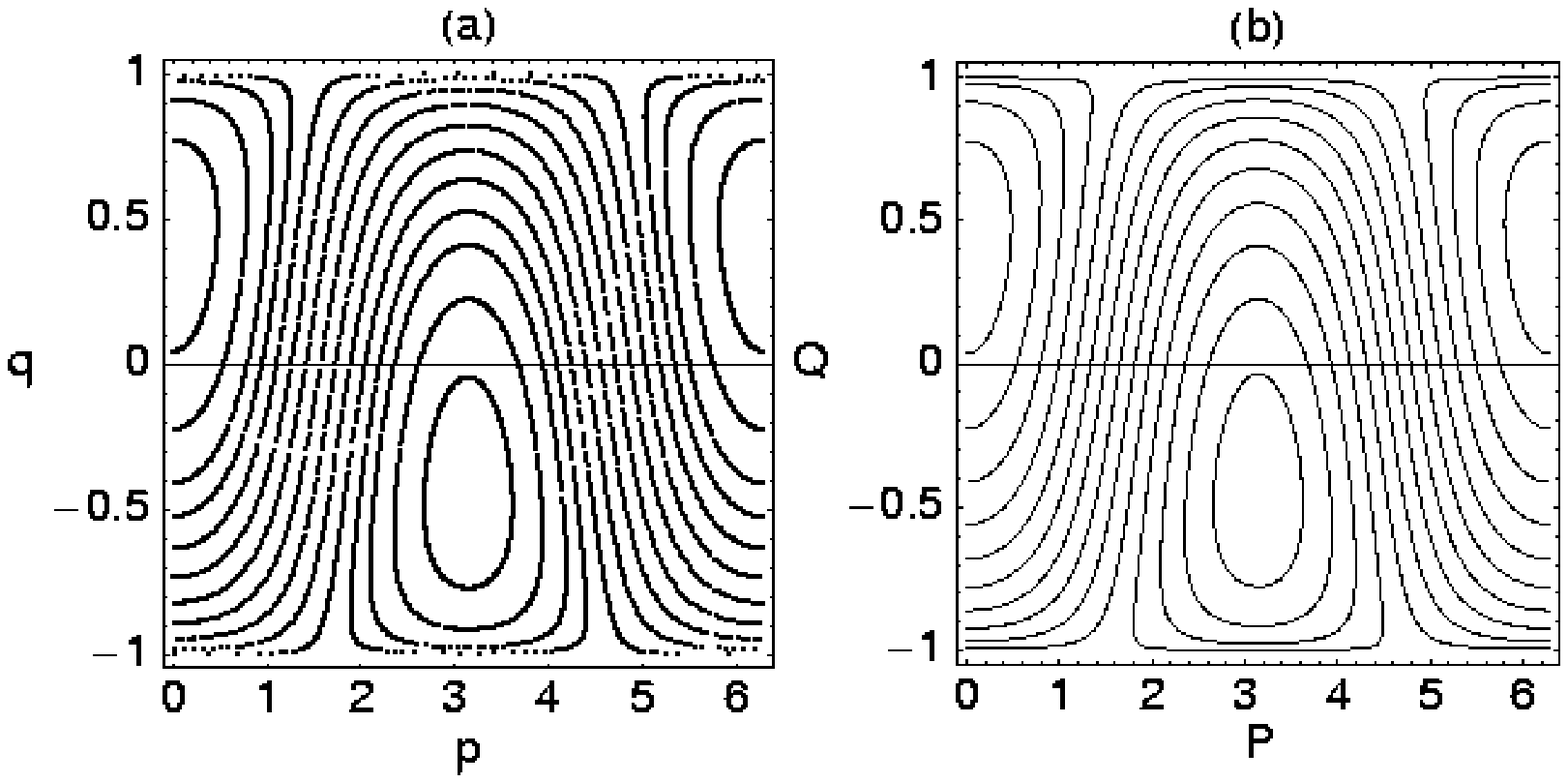} \\
\includegraphics[scale=0.45,angle=0]{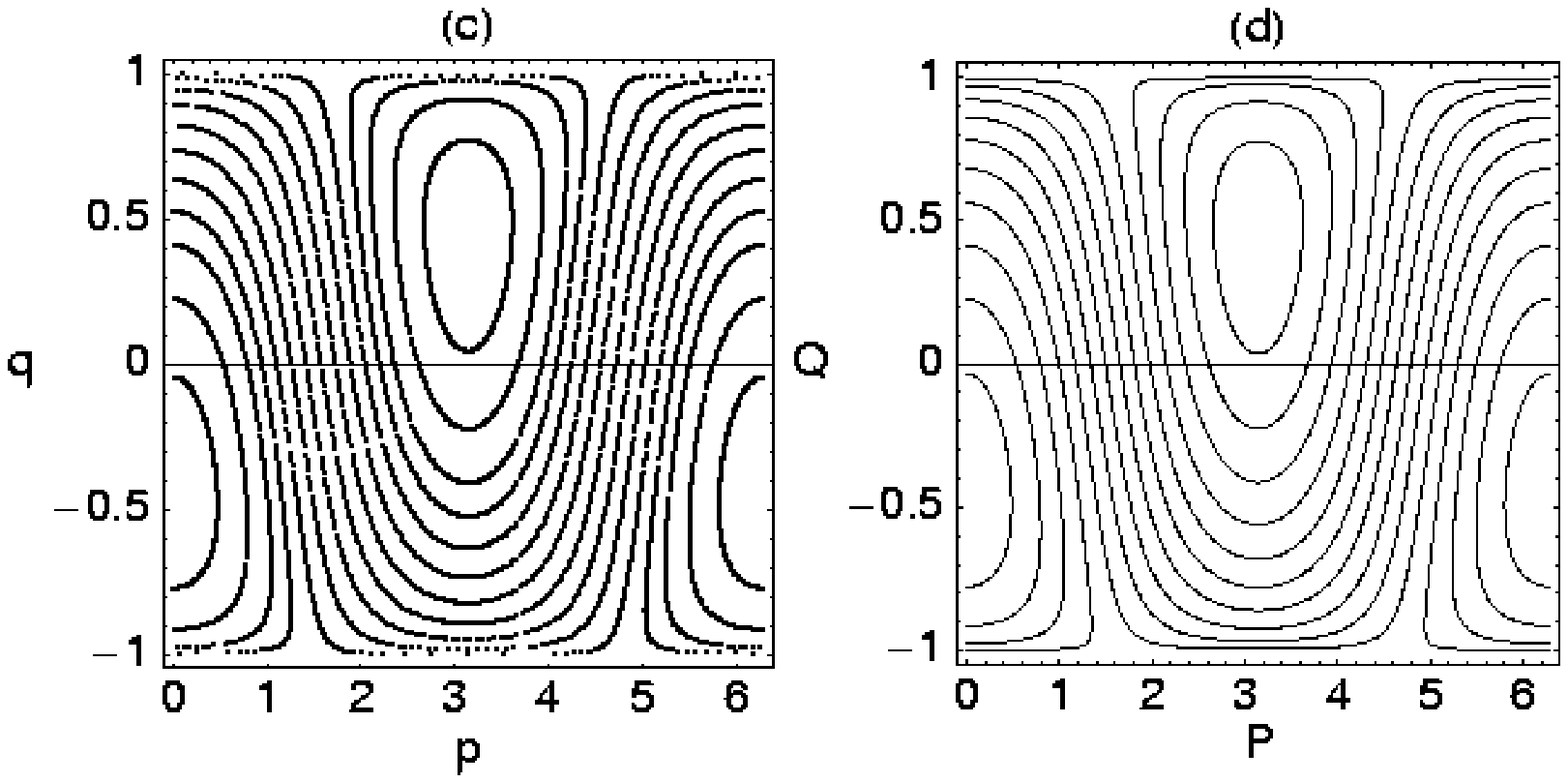} \\
\includegraphics[scale=0.45,angle=0]{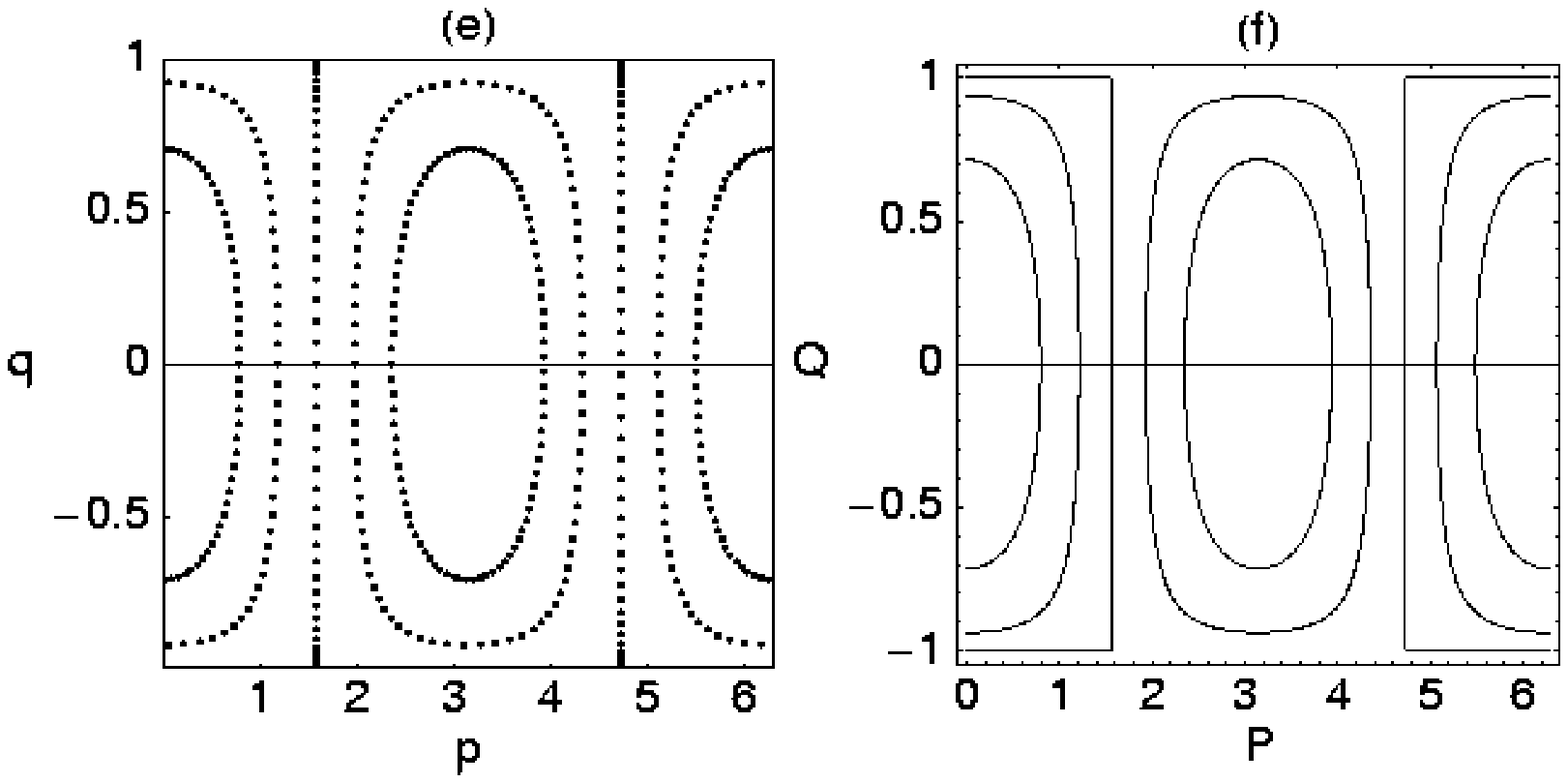}
\caption{\small Stroboscopic maps obtained numerically from the whole evolution (a,c,e) and respective contour curves obtained by \eqref{Smap} (b,d,f) for the R field with: $B_0=1$, $B_3=0$, $\omega=1$ and $\frac{B}{\omega}=\sqrt{5}$ (a,b), $B_0=B_3=\omega=0.5$ and $\frac{B}{\omega}=\sqrt{5}$ (c,d) and $B_0=1$, $B_3=\frac{\omega}{2}$, $w=89$ and $\frac{B}{\omega}=\frac{2}{89}$ (e,f). In (e) we see a case of rational tori. $\phi=0$ in all cases.}
\label{fig1}
\end{figure}

Since we have obtained the analytical solution for the system, the construction of the stroboscopic map is trivial. However, another procedure based on qualitative description of the stroboscopic map will be more appropriate for the comparative analysis that will be realized in the next sections. Re-writing \eqref{KH} at the stroboscopic instants one obtains
\begin{eqnarray}
\cal{K}=2 B_0\sqrt{1-q_k²}\cos{(p_k+\phi)}-2\,\left(B_3-\frac{\omega}{2} \right)\,q_k,
\label{Smap}
\end{eqnarray}
in which $r_k$ denotes $r(t_k)$. This formula allows one to construct the stroboscopic map by means of the contour curves. It is an alternative indication of the integrability of the system, since the map is composed only by tori (contour curves). This observation avoids (numerical) dynamical evolution, although it does not offer information about the comensurability pattern of the torus (rational or irrational), which may be obtained only by the analytical solutions. In fact, by means of the analytical equations of motion (see appendix \ref{app:analytical}) we found the conditions of comensurability, given by the ratio $\frac{B}{\omega}$, being $B=2\sqrt{B_0²+\Omega²}$ the amplitude of the magnetic field in the rotating frame (see appendix \ref{app:Gpicture}). Notice that rational and irrational tori cannot coexist in the same map, since the comensurability parameter, $\frac{B}{\omega}$, does not depend on the initial conditions. Figure \ref{fig1} illustrates these informations by some numerical examples.
\subsection{Periodic Orbits}\label{sec:po}
At stroboscopic instants both set of variables, $(p,q)$ and $(P,Q)$ yield the same map, since $Q_k=q_k$ and $P_k\pmod{2\pi}=p_k\pmod{2\pi}$. Then we can calculate the fixed points (periodic orbits of period one), denoted by $(\bar{P},\bar{Q})$, by imposing that $\dot{Q}=\dot{P}=0$ in the equations of Hamilton. This yields
\begin{subequations}
\begin{eqnarray}
\bar{P}&=&l \pi \qquad (l=0,1,2), \\ \nonumber \\
\bar{Q}_{\pm}&=&\mp\frac{2\Omega}{B}.
\end{eqnarray}
\label{po}
\end{subequations}
According to \eqref{psiqp} the corresponding quantum state (in the rotating frame) is
\begin{eqnarray}
|\varphi_{\pm}\rangle=\sqrt{\frac{1-\bar{Q}_{\pm}²}{2}}|+\rangle\pm\sqrt{\frac{1+\bar{Q}_{\pm}²}{2}}|-\rangle.\label{eigenket}
\end{eqnarray}
Putting $(\bar{P},\bar{Q})$ in the Hamiltonian (\ref{K}), we obtain  the contour value $\cal{K}=\bar{E}$, to which the fixed point belongs:
\begin{eqnarray}
\bar{E}=\pm B.\label{eigenvalue}
\end{eqnarray}
Equations \eqref{eigenket} and \eqref{eigenvalue} given the eigenstates and the eigenvalues of the quantum Hamiltonian associated to $\cal{K}$. Actually the eigenvalues are $\pm B/2$, as one may guess by \eqref{HQ} and \eqref{HC}. This is a very interesting example of the correspondence between eigenstates and periodic orbits. Another interesting point is that the frequency $\Omega=B_3-\frac{\omega}{2}$ controls the vertical localization of the fixed point and, consequently, the symmetry of the map. For instance, in the resonance ($\Omega=0$) one obtains $\bar{Q}=0$, as can be seen in Fig. \ref{fig1}.

Finally we point out that initial conditions like \eqref{po} are fixed point for all times in the phase space associated to variables $(P,Q)$. However, this is not so for the original variables, since that by \eqref{F3} we have  $(p(t),q(t))=(\bar{P}-\phi+\omega t,\bar{Q}_{\pm})$. In fact, in these phase space the fixed point occurs only at stroboscopic instants.
\section{The Nonrotating Field}\label{sec:NR}
We are now interested in analyzing the {\em nonrotating} (NR) field defined by \cite{bagrov01}
\begin{eqnarray}
\mathbf{B}_{NR}=-2(B_0,0,B_3 \cos{\omega t}),\label{BNR}
\end{eqnarray}
being $B_0$ and $B_3$ constant amplitudes. Actually this is one of the most common fields used in NMR experiments, mainly for quantum computation purposes \cite{divicenzo95}, even more because fields like \eqref{BR} are difficult to manufacturate. The corresponding Hamiltonian reads
\begin{eqnarray}
\cal{H}(q,p,t)=2 B_0\sqrt{1-q²} \cos{p}-2 B_3\, q \cos{\omega t}.
\label{HNS}
\end{eqnarray}
The absence of a rotation symmetry destroys the manifest integrability of the model. In the R system, $B_R=||\mathbf{B}_R||$ is constant, so that the time variation of the external field is given only by changes in its direction. Then one may choose a frame that rotates with the field frequency, getting a constant energy. Because in the NR system the field varies also in modulus it is not possible to eliminate time-dependence just by defining a rotating frame. In fact we are not able to find a canonical transformation that puts the Hamiltonian in an autonomous form. The quantum version of this system has been analyzed in different contexts, but always in a perturbative way \cite{barata00,wresz98,frasca03}.
\subsection{Numerical Analysis}\label{sec:numerical}
In this section we present some numerical results concerning the features of the stroboscopic map and its qualitatively ressemblance with those of the R system. Consequently, integrability is numerically verified.

\begin{figure}[ht]
\centerline{\includegraphics[scale=0.3,angle=-90]{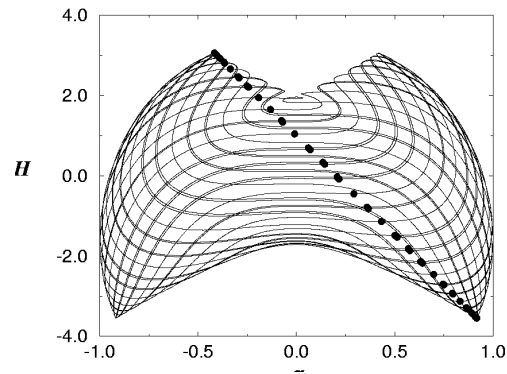}}
\caption{\small Parametric plot between the time-dependent energy $\cal{H}(t)$
and the coordinate $q(t)$ for the continuous flux (solid line) and
for the stroboscopic instants $t_k$ (bullets).The parameters used
in this calculation were $B_0=1.0$, $B_3=1.5$ and $\omega=3.0$. }
\label{Eq}
\end{figure}

The first interesting information that emerges from the numerical results is the linear relation between the time-dependent energy, $\cal{H}(q(t),p(t),t)$, and the coordinate $q(t)$ at the stroboscopic instants $t_k=\frac{2\pi}{\omega} k$. In Fig. \ref{Eq} we illustrate this information by means of stroboscopic bullets. It is also possible to notice, by the continuous flux, the complexity of the NR dynamics (compare to \ref{EqS}). 
\begin{figure}[ht]
\centerline{\includegraphics[scale=0.3,angle=-90]{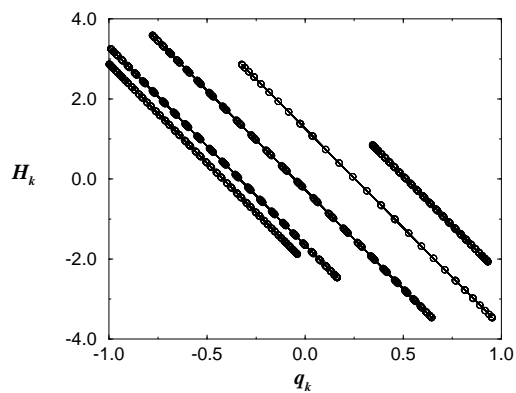}}
\caption{\small ``Stroboscopic parametric plot'' of $\cal{H}$ and $q$ (symbols) for several initial conditions and respective fittings (solid lines). The slopes of the fitting curves ($\gamma=4.9559$) are identical and independent of the initial condition. The parameters used in this calculation were $B_0=1.0$, $B_3=1.5$ and $\omega=3.0$.}
\label{fitting}
\end{figure}

In Fig. \ref{fitting} we show the same type of stroboscopic parametric plot for several initial conditions under the same parameter interaction. It is possible to see that the slope (regression coefficient), $\gamma$, of each straight line has always the same value, i.e, it does not depend on the initial condition. On the other hand, the regression constant, $\cal{E}$, does depend on the initial condition. 

These numerical results allow us to write $\frac{\Delta\cal{H}}{\Delta q} = -\gamma$ and
\begin{eqnarray}
\cal{H}_k = \cal{E}(q_0,p_0)-\gamma q_k, \label{EqNS}
\end{eqnarray}
in which $\Delta\cal{H}=\cal{H}_{k+l}-\cal{H}_k$, $\Delta q=q_{k+l}-q_k$ and $\cal{H}_k=\cal{H}(q_k,p_k,t_k)$. Using \eqref{HNS} we re-write equation above as
\begin{eqnarray}
\cal{E}(q_0,p_0) = 2 B_0 \sqrt{1-q_k²}\cos{p_k}-2\left(B_3-\frac{\gamma}{2} \right) q_k,
\label{NSmap}
\end{eqnarray}
which is the equation of the stroboscopic map for the NR system as a function of the numerical parameter $\gamma$.
Comparison between \eqref{EqNS} and \eqref{EqS} emphasizes the resemblance between NR and R dynamics, at least at the stroboscopic instants. 

\begin{figure}[ht]
\begin{center}
\centerline{\includegraphics[scale=0.3,angle=-90]{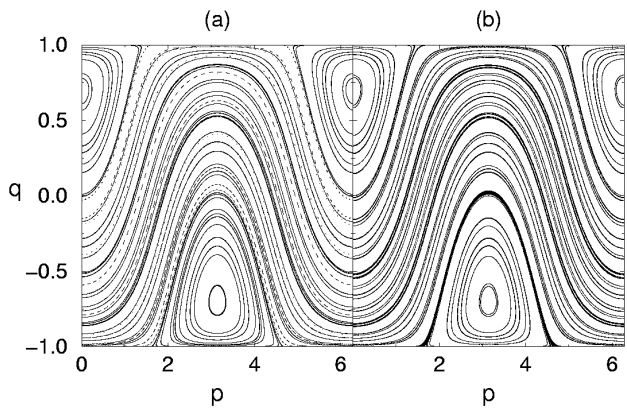}}
\end{center}
\caption{\small (a) Stroboscopic map for the NR system with $B_0=1.0$, $B_3=1.5$ and $\omega=3.0$ and (b) the corresponding map for the the R system with $\omega=\gamma=4.9559$ (see eq. \ref{Smap}). This value was obtained by the fitting shown in Fig. \ref{fitting}.}
\label{sec1}
\end{figure}

On the other hand, equation \eqref{NSmap} indicates that the maps of the NR system must be ``identical in shape'' to those of the R case, as can be seen by comparing \eqref{NSmap} and \eqref{Smap} (with $\phi=0$). The difference is only in the resonance condition, which is now given by $B_3-\frac{\gamma}{2}$, instead of $B_3-\frac{\omega}{2}$. $\gamma$ is a frequency determined numerically by the fittings shown in Fig. \ref{fitting}, whereas $\omega$ is the own frequency of the applied field. The qualitative agreement between NR and R maps are shown in Fig. \ref{sec1}, which was made as follows: we determined $\gamma$ numerically by fittings in the NR system and then we used a R field with $\omega=\gamma$.
Another interesting difference can be noticed in these numerical results: in the NR system, rational and irrational tori are allowed to coexist.

 The above results explain the fact observed in \cite{barata00} that the RWA accounts for several results qualitatively even outside the regime where it is supposed to be a good approximation \cite{nus73}, namely: i.) resonance and ii.) weak coupling. In particular the solution found in \cite{barata00} in the case of resonance has a frequency which is very close to the Rabi frequency.

In the next sections we demonstrate the explicit integrability (i.e.,the possibility of obtaining explicit analytical solutions) of the NR system at some regimes of interest.
\subsection{Analytical Analysis: Average Dynamics} \label{sec:analytical}
The contour pattern \eqref{NSmap} chosen by the NR system is not obvious. In fact it can not be inferred directly from the Hamiltonian \eqref{HNS}. The choice by one specific contour in the set of possibilities offered by the time-dependence of the system suggests the existence of an average dynamics. In this section we try to obtain some analytical evidence of such an idea.

Consider the following arbitrary Hamiltonian
\begin{eqnarray}
\cal{H}(q,p,t)=\cal{H}_0(q,p)+f(\omega t)V(q,p),
\label{Hg}
\end{eqnarray}
in which $f(\omega t)$ is a periodic function with period $T=\frac{2\pi}{\omega}$. We now calculate the time-derivative of the Hamiltonian,
\begin{eqnarray}
\dot{\cal{H}}&=& V\partial_t f = \partial_t\left(Vf\right) - f \partial_t V,
\label{Hdot}
\end{eqnarray}
and then integrate the result over the arbitrary interval $[t_1,t_2]$ to get a time-dependent function given by
\begin{eqnarray}
\cal{H}_2-\cal{H}_1=f_2V_2-f_1 V_1-
\int_{t_1}^{t_2} f(\omega t) \dot{V}(t) dt, \label{Ht}
\end{eqnarray}
in which we used the notation $\cal{A}_i=\cal{A}(t_i)$, with $\cal{A}$ standing for $\cal{H}$, $V$ and $f$. Now, assuming that $V(t)$ is not a constant of motion, we multiply and divide the last term of the equation above by $V_2-V_1=\int_{t_1}^{t_2} \dot{V}(t)dt$ and reorganize the equation in the form
\begin{eqnarray}
\cal{H}_2-\cal{H}_1=\Big(f_2-\langle f \rangle_{2,1}\Big)V_2 -\Big(f_1-\langle f \rangle_{2,1}\Big)V_1  \label{intdotH}
\end{eqnarray}
in which we have defined the average
\begin{eqnarray}
\langle f \rangle_{2,1} \equiv \frac{\int\limits_{t_1}^{t_2} f(\omega t)\dot{V}(t)dt}{\int\limits_{t_1}^{t_2}\dot{V}(t)\,dt}.\label{f21}
\end{eqnarray}
This equation can be re-write as
\begin{eqnarray}
 \int\limits_{t_1}^{t_2} f(\omega t)\dot{V}(t)dt =\langle f \rangle_{2,1}\Big(V_2-V_1 \Big),\label{fDV}
\end{eqnarray}
which presumes that the integral of $f\dot{V}$ may be identified to the product of the integral of $\dot{V}$ alone multiplied by a certain average.

At arbitrary stroboscopic instants, namely $t_1=i T$ and $t_2=j T$, with $i$ and $j$ $\in \mathbb{N},$ \eqref{f21} and \eqref{fDV} may be sufficiently manipulated to yield
\begin{eqnarray}
\langle f \rangle_{j,i} = \frac{\int\limits_0^{T} f(\omega t)\sum\limits_{n=i}^{j-1}\dot{V}(t+nT)\,dt}{\int\limits_0^{T}\sum\limits_{n=i}^{j-1}\dot{V}(t+nT)\,dt}\label{fji}
\end{eqnarray}
and 
\begin{eqnarray}
\Big(V_{j}-V_{i}\Big)\langle f \rangle_{j,i} = \sum\limits_{n=i}^{j-1}\Big(V_{n+1}-V_n \Big)\langle f \rangle_{n+1,n}. \label{fDVu}
\end{eqnarray}
Convenient transformations of variable were performed in order to put the formulas in terms of integrals over only one period. 

Consider the behavior of the mean \eqref{fji} at consecutive stroboscopic instants for long times $(k\gg 1)$. In such case we have
\begin{eqnarray}
\langle f \rangle_{k+1,k}=\frac{\int\limits_0^T f(\omega t)\dot{V}(t+kT)dt}{\int\limits_0^T\dot{V}(t+kT)dt}. \label{fkk}
\end{eqnarray}
For any value of $T$ there exist a value of $k$ large enough to guarantee that $kT\gg t$. It allows us to expand $\dot{V}(kT+t)$ in Taylor's series, being $t$ the small parameter. This procedure leads to
\begin{eqnarray}
\langle f \rangle_{k+1,k}=\frac{\sum\limits_{n=0}^{\infty}\frac{\partial_{t_k}^n \dot{V}(t_k)}{n!}\left[\frac{1}{\omega^{n+1}}\int\limits_0^{2\pi} f(\varphi)\varphi^nd\varphi\right]}{\sum\limits_{n=0}^{\infty}\frac{\partial_{t_k}^n \dot{V}(t_k)}{n!}\left[\frac{1}{n+1}\left(\frac{2\pi}{\omega}\right)^{n+1} \right]},\label{expand}
\end{eqnarray}
in which $\varphi=\omega t$. In this result all dependence with $\omega$ was made explicit. Then, the regime of high frequency becomes evident for well-defined functions $f$ and $\dot{V}$: 
\begin{eqnarray}
\lim\limits_{\omega\to\infty \atop (k\gg 1)}\langle f \rangle_{k+1,k} = \frac{1}{T}\int\limits_{0}^T f(\omega t) dt\equiv \bar{f}.
\end{eqnarray}
In such situation, according to \eqref{fDVu}, we have $\langle f\rangle_{j,i}=\bar{f}$. Then, choosing $t_2=t_k$, $t_1=0$ and putting results above in \eqref{intdotH} we obtain
\begin{eqnarray}
\cal{H}_k=\Big[\cal{H}_0(q_0,p_0)+\bar{f} V(q_0,p_0)\Big]+\Big[f_0-\bar{f}\Big] V_k, \label{HkNS}
\end{eqnarray}
or
\begin{eqnarray}
\cal{E}(q_0,p_0)=\cal{H}_k-\Big[f_0-\bar{f}\Big] V_k, \label{NSmap-w}
\end{eqnarray}
with $\cal{E}(q_0,p_0)=\cal{H}_0(q_0,p_0)+\bar{f} V(q_0,p_0)$. This result attests the integrability of the system, since it provides the analytical expression for the stroboscopic map. Notice that
\begin{eqnarray}
\frac{\Delta \cal{H}}{\Delta V}=f_0-\bar{f}, \label{DHDVNS}
\end{eqnarray}
and thus \eqref{HkNS} and \eqref{DHDVNS} can be directly compared to respective results \eqref{EqS} and \eqref{dHdq} for the integrable system R. Once the system is notably integrable for $k\gg 1$ it is expected that it is also true for every $k$. It must be so to guarantee that the map is composed by the same contours at any stroboscopic instant. 

The regime of low frequency can not be predicted directly from the formulas developed in this section and numerical work is necessary. However, the high frequency regime illustrates the necessary and sufficient condition for the integrability of the system \eqref{Hg}:
\begin{eqnarray}
\langle f\rangle_{k+1,k}=\langle f\rangle. \label{condition}
\end{eqnarray}
This condition requires that the average defined by \eqref{fkk} be time-independent. One further observation should be done in order to emphasize that the average dynamics is weighted up by the time-derivative of the potential and not only by the time-average of the function $f$. This fact credits a non-trivial character to the identification of such an average.

In the next section, we obtain some explicit regime of integrability by applying the average we defined above, the averaging theorem and the RWA.

\subsection{\bf Analytical Solutions in the NR System}
\subsubsection{\bf High Frequency Regime}\label{sec:high}
In this section we consider the regime of high frequency, in which the frequency of the external field $\omega$ is much larger than $B_0$ or $B_3$. In fact, we will work in a regime that allows us to disregard higher order terms in \eqref{expand}.

\begin{figure}[ht]
\centerline{\includegraphics[scale=0.3,angle=-90]{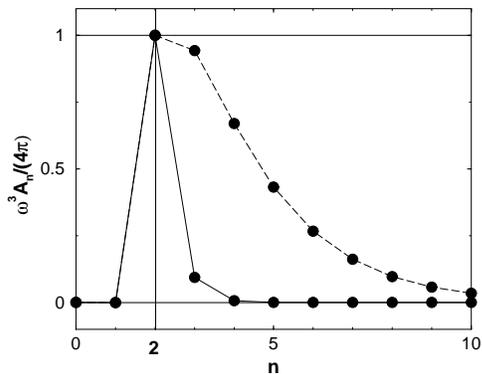}}
\caption{\small $\frac{\omega^3 A_n}{4\pi}$ as a function of integers $n$ for $\omega=10$ (dashed line) and $\omega=100$ (solid line), in arbitrary units. The maximum value of the function plotted does not depend on $\omega$.}
\label{An}
\end{figure}

At first we investigate the terms 
\begin{subequations}
\begin{eqnarray}
A_n\equiv \frac{1}{\omega^{n+1}}\int_{0}^{2\pi}f(\varphi)\varphi^n d\varphi
\end{eqnarray}
and
\begin{eqnarray}
B_n\equiv \frac{1}{n+1}\left(\frac{2\pi}{\omega}\right)^{n+1},
\end{eqnarray}
\end{subequations}
 that appear in the brackets in \eqref{expand}. For $f(\varphi)=\cos{\varphi}$ the integral above can be performed analytically, leading to a hyper-geometric function with a maximum at $n=2$. In Fig. \ref{An} we show the behavior of $\omega^3 A_n/4\pi$ as a function of $n$ for two values of $\omega$.
For high frequency regimes we indeed verify that only $n=2$ contributes significantly to the sum, i.e., only $A_2$ is non-null, though small. On the other hand, for $B_n$, the major contribution comes from $n=0$, as it is evident from its own definition.

According to Fig. \ref{An}, we can write $A_n=\frac{4\pi}{\omega^3}\delta_{n,2}$. Then, \eqref{expand} assumes the form
\begin{eqnarray}
\langle f\rangle_{k+1,k}=\frac{1}{\omega²}\frac{\partial_{t_k}^2\dot{V}(t_k)}{\dot{V}(t_k)}.\label{fk}
\end{eqnarray}
Now, by noticing that $\dot{V}=-2 B_3\dot{q}$ and using the equations of motion of the NR system, we get
\begin{eqnarray}
\langle f\rangle_{k+1,k}=-4\left(\frac{B_0²+B_3²}{\omega²}\right)\equiv \langle f\rangle,\label{<f>}
\end{eqnarray}
which is notably $k$-independent. This result explicits the condition of high frequency that we supposed initially.

\begin{figure}[ht]
\centerline{\includegraphics[scale=0.3,angle=-90]{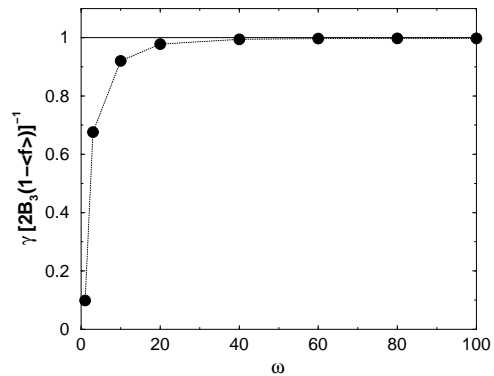}}
\caption{\small. Comparison between the numerical $\gamma$ and analytical $2B_3(1-\langle f \rangle)$ slopes for several values of $\omega$.}
\label{gammas}
\end{figure}

For large values of $\omega$ (not infinity) we can replace $\bar{f}$ in \eqref{HkNS} by the result \eqref{<f>}. Then, by comparing \eqref{HkNS} with \eqref{EqNS}, we establish the connection between the analytical ($\langle f \rangle$) and numerical ($\gamma$) results as follows
\begin{eqnarray}
\gamma=2B_3\left(1-\langle f \rangle \right).\label{gammaf}
\end{eqnarray}
In Fig. \ref{gammas}, the validity of this relation for high frequencies is verified, showing that our approximations are adequate. 

The analytical expression for the stroboscopic map in the regime of high frequency can be constructed by putting $V_k=-2B_3q_k$ in \eqref{HkNS} with the explicit form of $\cal{H}_k$:
\begin{eqnarray}
\cal{E}(q_0,p_0)=2B_0\sqrt{1-q_k²}\cos{p_k}-2B_3\langle f\rangle q_k,\label{NSmap-<f>}
\end{eqnarray}
with $\langle f \rangle$ given by \eqref{<f>}. Notice that $\gamma\to 2B_3$ and $\langle f\rangle\to 0$ in the limit of $\omega\to\infty$. It shows that the map will have periodic orbits localized in the line $q=0$. In this limit, accordingly to both \eqref{NSmap} and \eqref{NSmap-<f>}, the stroboscopic map reads
\begin{eqnarray}
\cal{E}(q_0,p_0)=2B_0\sqrt{1-q_k²}\cos{p_k}.\label{NSmap-w0}
\end{eqnarray}

The approach adopted in this section allows us to predict the form of the stroboscopic map in the regime of high frequency, defined precisely by \eqref{<f>}. However this cannot be achieved for any periodic function $f$. In fact, it is possible to show that the average is $k$-dependent for $f(t)=\sin{\omega t}$.
\subsubsection{\bf Strong Coupling Regime} \label{sec:strong}
In order to analyze the strong coupling regime ($B_0\ll \omega$) we apply the canonical transformation \eqref{F3} but now with $\theta=\frac{2B_3}{\omega}\sin{\omega t}$ (and $\phi=0$). The new Hamiltonian reads
\begin{eqnarray}
\cal{K}&=& 2 B_0 \sqrt{1-Q²}\cos{\left(P+\frac{2 B_3}{\omega}
\sin{\omega t} \right)}\label{Kstrong}
\end{eqnarray}
and the equations of motions are given by
\begin{subequations}
\begin{eqnarray}
\dot{Q}&=&-2B_0\sqrt{1-Q²}\sin{\left(P+\frac{2 B_3}{\omega}\sin{\omega t} \right)},\\ \dot{P}&=&2B_0\frac{Q}{\sqrt{1-Q²}}\cos{\left(P+
\frac{2 B_3}{\omega}\sin{\omega t} \right)}.
\end{eqnarray}\label{dQP}
\end{subequations}
If $B_0 T$ is small these equations will be in the form
\begin{eqnarray}
\dot{x}=\varepsilon f(x,\omega t),
\end{eqnarray}
with $\varepsilon\,T\ll 1$, and the averaging theorem can be applied 
\cite{verhulst90}.
The mean solution, $x_m$, satisfying $|x-x_m|< c\varepsilon$, being $c$
independent of $\varepsilon$, for $t \in [0,\frac{1}{\varepsilon}]$,
will be given by
\begin{subequations}
\begin{eqnarray}
&\dot{x}_m=\varepsilon f^0(x_m),&  \\
&f^0(x)=\frac{1}{T}\int\limits_0^T f(x,\omega t) dt,& \label{average}
\end{eqnarray}
\end{subequations}
where $T=\frac{2\pi}{\omega}$ is the period defined by $f(x,\omega 
t+\omega T)=f(x,\omega t)$.

By applying this scheme to the equations of motion (\ref{dQP}) we get
\begin{subequations}
\begin{eqnarray}
\dot{Q}_m&=&-2\,\omega_0\sqrt{1-Q_m²}\sin{P_m},\\
\dot{P}_m&=&2\,\omega_0\frac{Q_m}{\sqrt{1-Q_m²}}\cos{P_m},
\end{eqnarray} \label{dQPm}
\end{subequations}
being 
\begin{eqnarray}
\omega_0=B_0\,J_0\left(\frac{2B_3}{\omega} \right).\label{w0}
\end{eqnarray}
$J_0(x)$ is the Bessel function of first kind. Notice that we may have dynamical localization ($\omega_0=0$) for values of $B_3$ and $\omega$ that produce zeros in $J_0$ \cite{barata03,frasca03loc}.

Equations above were obtained by expanding the (co)sine in terms of Bessel functions sums. The system (\ref{dQPm}) can be solved analytically by noticing that
\begin{eqnarray}
\cal{K}_m\equiv 2 \omega_0 \sqrt{1-Q_m²}\cos{P_m} \label{Km}
\end{eqnarray}
is a constant of motion. Note that this result can be obtained by applying the formula \eqref{average} to Hamiltonian \eqref{Kstrong}. Then, since by the averaging theorem $Q(t)\approx Q_m$, $P(t)\approx P_m$ and $\cal{K}(t)\approx\cal{K}_m$ for $t\le\frac{1}{B_0}$, we can return to the original set of variables to obtain the analytical solution. However, for our purpose, its sufficient to establish the connection between the original and the new Hamiltonian at the stroboscopic times. For all $t$ satisfying the condition of validity of the averaging theorem, namely $B_0 T = \frac{2\pi B_0}{\omega}\ll 1$, we have $\cal{K}_m=\cal{H}+2B_3 q \cos{\omega t}$, which allows us to conclude that
\begin{eqnarray}
\cal{H}_k=\cal{K}_m-2 B_3 q_k.
\end{eqnarray}
This equation allows us to obtain the analytical expression for the stroboscopic map in the strong coupling regime:
\begin{eqnarray}
\cal{K}_m=2\omega_0\sqrt{1-q_k²}\cos{p_k},\label{NSmap-strong}
\end{eqnarray}
which could had been obtained from \eqref{Km} by noticing that $(P_m(t_k),P_m(t_k))=(P_k,Q_k)=(p_k,q_k)$.

Direct comparison with \eqref{EqNS} and \eqref{gammaf} indicates that this regime is equivalent to that of high frequency, since here we have $\gamma=2B_3$ and $\langle f\rangle=0$. Then, once more the map will be completely symmetric in $q$ and $p$, with central periodic orbits, as those of the Fig. \ref{fig1}-(e,f). The strong coupling regime has been studied in a more general case (quasi-periodic fields) in \cite{wresz98}.

It is also worth to observe that the averaging theorem must be valid for all times and not only until $t_{max}=\frac{1}{B_0}$. If not, the map would mix two types of contours, one determined by $\langle f \rangle=0$ and other by a general $\langle f\rangle \neq 0$. In this case, tori crossing would be inevitable.

\begin{figure}[ht]
\centerline{\includegraphics[scale=0.3,angle=-90]{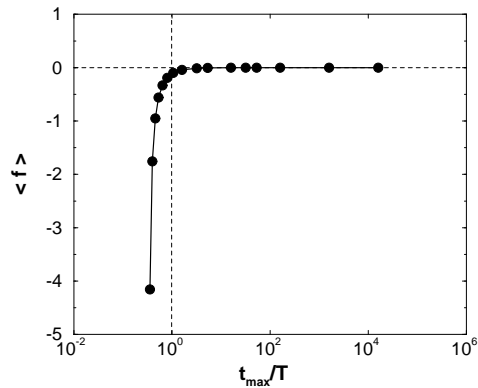}}
\caption{\small The average weighted up by the potential time-rate,
$\langle f\rangle$, as a function of $\frac{t_{max}}{T}= \frac{1}{B_0 T}=\frac{1}{2\pi}\frac{\omega}{B_0}$, the maximal normalized instant for the which is valid the average theorem solution. In this simulation we fixed $B_3=1.0$, $\omega=10$ and made vary $B_0$. According to the averaging theorem $\langle f\rangle=0$.}
\label{fig:fk}
\end{figure}

We emphasize this point by means of the numerical calculation shown in Fig. \ref{fig:fk}. The plot shows the time-independent average $\langle f\rangle$, calculated numerically from the whole evolution (valid for all time), as a function of $\frac{t_{max}}{T}=\frac{1}{B_0 T}=\frac{1}{2\pi}\frac{\omega}{B_0}$. For small values of $t_{max}/T$ we have $\langle f\rangle \neq 0$, but at such regime the averaging theorem is no longer applicable. In fact, according to the numerical results $\langle f\rangle$ already differs from zero for $t_{max}/T\approx 1.6$, which corresponds to $B_0 T \approx 0.6$, whereas the theorem is valid only for $B_0 T \ll 1$. Consider now the region of large values of $t_{max}/T$. We see that $\langle f\rangle$ is null, in agreement with the averaging theorem. Since $\langle f\rangle$ is the value valid for all times (and not only until $t_{max}$) we may conclude that the averaging theorem, proved valid for large times, must be valid for all times.
\subsubsection{\bf Resonant Weak Coupling Regime: RWA}\label{sec:rwa}
In this section we apply the averaging theorem to treat the NR system in a regime defined by both a resonance condition and one small parameter. These two assumptions are essential to the application of the RWA. Actually our aim is showing that the averaging theorem ensures the validity of the RWA.

Consider the Scrödinger equation corresponding to $\mathbf{B}_{NR}$. After a time independent rotation of $\pi/2$ around the $y$-axis we obtain
\begin{eqnarray}
\imath\,\frac{d\psi}{dt}=-2\Big(B_0\,\sigma_z-B_3\cos{\omega t}\,\sigma_x \Big)\psi.\label{psiy}
\end{eqnarray}
We now perform a unitary transformation to the rotating frame, 
\begin{subequations}
\begin{eqnarray}
\psi(t)=R(t)\psi'(t), 
\end{eqnarray}
with
\begin{eqnarray}
R(t)=\exp{\left(-\imath\, \omega t \sigma_z/2\right)},
\end{eqnarray}\label{R}
\end{subequations}
obtaining
\begin{eqnarray}
\imath\,\frac{d\psi'}{dt}=\left[\left(B_0-\frac{\omega}{2} \right)\sigma_z-B_3\, \cos{\omega t} \,\Big(R^{-1} \sigma_x R\Big)\right]\,\psi'.\label{dtpsil}
\end{eqnarray}
Now,
\begin{eqnarray}
\cos{\omega t} \,R^{-1} \sigma_x R=\frac{\sigma_x}{2}+\frac{1}{2}\Big(\sigma_x \cos{2\omega t}-\sigma_y \sin{2\omega t} \Big)\label{high}.
\end{eqnarray}
The second term at the r.h.s. of \eqref{high} produces the high frequency oscillations. According to the heuristics \cite{baym}, they may be neglected for the purposes of ascertaining the ``mean'' behaviour in time. We now show the conditions under which this is true. Assume that
\begin{subequations}
\begin{eqnarray}
\omega=2 B_0 \qquad \,\,\,\,\, \textrm{(resonance)}\,\,\,\,\,\,\label{resso}
\end{eqnarray}
and
\begin{eqnarray}
B_3/\omega \ll 1 \qquad \textrm{(weak coupling).}\label{wc}
\end{eqnarray}\label{RWAconditions}
\end{subequations}
Putting \eqref{high} into \eqref{dtpsil} we obtain
\begin{eqnarray}
\imath \frac{d\psi'}{dt}=-\frac{B_3}{2}\,\Big(\sigma_x + \sigma_x\cos{2\omega t}-\sigma_y \sin{2\omega t}\Big)\psi'.\label{psilresult}
\end{eqnarray}
Under assumption \eqref{wc}, it follows, from the averaging theorem \cite{verhulst90} (Theorem 11.2, pg. 154):
\begin{eqnarray}
\left\|\psi'(t)-\psi_0'(t) \right\|=\cal{O}\left(B_3/\omega \right)\label{norm}
\end{eqnarray}
in the time-scale $\cal{O}\left(\omega/B_3 \right)$, i.e., for all $t\in\left[0,\frac{c}{(B_3/\omega)}\right]$, where $c$ is a constant, $\|\cdot\|$ is the Euclidean norm in spinor space and $\psi_0'(t)$ satisfies the ``averaged equation''
\begin{eqnarray}
\imath\frac{d\psi_0'}{dt}=-\frac{B_3}{2}\sigma_x\,\psi_0'.\label{aver}
\end{eqnarray}
From \eqref{R}, \eqref{norm} and \eqref{aver} we see that
\begin{eqnarray}
\left\|\psi(t)-\psi_{RWA}(t) \right\|=\cal{O}\left(B_3/\omega \right), \label{normRWA}
\end{eqnarray}
for $t\in\left[0,\frac{c}{B_3/\omega}\right]$, where
\begin{eqnarray}
\psi_{RWA}(t)\equiv R(t)\,e^{\imath B_3 t \sigma_x/2}\,\psi(0).
\end{eqnarray}
It is possible to estimate the effect of being slightly off resonance by \cite{verhulst90} (Theorem 11.2, pg. 154). By \eqref{dtpsil} and \eqref{high}, defining the Rabi frequency
\begin{eqnarray}
\Omega_R=\left[\left(2 B_0-\omega \right)²+B_3² \right]^{1/2},
\end{eqnarray}
then $\psi'$ given by \eqref{psilresult} is close (in the sense of \eqref{normRWA}) to the solution 
\begin{eqnarray}
\psi'(t)=e^{-\imath\, \Omega_R t\,\, \sigma/2}\,\psi'(0),
\end{eqnarray}
where
\begin{eqnarray}
\sigma=\frac{(2 B_0-\omega)}{\Omega_R}\,\sigma_z-\frac{B_3}{\Omega_R}\,\sigma_x,
\end{eqnarray}
with $\sigma²=1$. It is to be remarked that an exact method suggested by the RWA was devised in \cite{jauslin}. This powerful method (called in \cite{jauslin} rotating wave transformation) goes well beyond the simple idea of the RWA, and has recently been applied to the quantum case \eqref{HFQ} in \cite{amniat}. We have been concerned, in this section, with the standard problem of justifying the replacement of $\mathbf{B}_{R}$, given by \eqref{BR}, by $\mathbf{B}_{NR}$, given by \eqref{BNR}, for which, surprisingly, no precise estimates on the parameters seen to have been given.

\section{Unitary Quantum NOT Operation} \label{NOTgate}

In this section we focus on the potential capability of the two-level systems in working as quantum NOT gates. We are interested mainly in determining the conditions for the realization of the NOT operation, pointing out the alternative regimes for its experimental implementation.

General principles of quantum computation require that NOT operation be unitary \cite{gisin99}. References \cite{gisin99,buzek99} propose different constructions of the unitary NOT operation, e.g., working in an extended Hilbert space. Although these formalisms are also appicable to our case, we are, primarily, concerned with the dynamics We thus choose to keep the original setting, in particular, the same Hilbert space. In this case, unitarity of the NOT operation may restrict us to special cases of initial superpositions. It indeed occurs, as we will see.

Our analysis starts by the analytical study of the rotating field treated in section \ref{sec:R}, for the which all analytical solutions are available (appendix \ref{app:analytical}). Since a resonant NR field in the regime of weak coupling can be mapped on a system governed by a rotating field, as showed in the last section, this case will be automatically included in the analysis.

The unitary quantum NOT operation is characterized by the transformation $\mathbf{S}(0)\to\mathbf{S}(t_{not})$ in the classical phase space. This transition, achieved by means of a unitary Hamiltonian dynamics, leads the initial vector $\mathbf{S}(0)$ to its antipode $\mathbf{S}(t_{not})$ in the unit sphere. $t_{not}$, defined by $\mathbf{S}(t_{not})\cdot\mathbf{S}(0)=\mathbf{S}_{not}\cdot\mathbf{S}_0=-1$, is the instant at which the NOT operation occurs. 

Thus, imposing $\mathbf{S}_{not}\cdot\mathbf{S}_0=-1$ in the analytical result given in \ref{SSrwa} we find several regimes of operation for the NOT-gate. Bellow we list all of them.

\vspace{0.3cm}
\noindent
{\bf CASE 1} \\
Satisfied the resonance condition
\begin{subequations}
\begin{eqnarray}
\omega²=B_0²+\left(B_3-\frac{\omega}{2} \right)²,
\end{eqnarray}
which corresponds to $\omega=\frac{B}{2}$, the NOT operation will occur periodically at
\begin{eqnarray}
t_{not}^{(n)}=(2n+1)\frac{\pi}{\omega} \qquad (n\in\mathbb{N})
\end{eqnarray}
for the following (equivalent) sets of initial conditions
\begin{eqnarray}
(p_0,q_0)&=&(\forall,0), \\
\mathbf{S}_0&=&(\cos p_0,\sin p_0,0), \\
|\psi_0\rangle&=&\frac{1}{\sqrt{2}}\Big(|+\rangle+e^{\imath p_0}|-\rangle \Big).
\end{eqnarray}
\end{subequations}
Note that the only participants of the unitary NOT operation are initial conditions belonging to the plane $xy$, specifically on the equatorial line of the unit sphere. This situation corresponds to a kind of {\em universal phase} NOT-gate, since all relative phase $p_0$ will be transformed into $p_0+\pi$ at $t_{not}$. We will refer to these solutions as ``$b\varphi$'', denoting the branch in which all $\varphi$ is valid (equatorial line). From now on we will use the notation ``$bp_0$'' to denote the {\em branch} of solutions corresponding to the initial conditions related to $p_0$. 

\vspace{0.3cm}
\noindent
{\bf CASE 2} \\ Under the resonance conditions 
\begin{subequations}
\begin{eqnarray}
B_3&=&\frac{\omega}{2}, \\
B_0&=&(2 m +1) \frac{\omega}{2} \qquad (m\in\mathbb{N}),
\end{eqnarray}
the NOT operation occurs periodically at
\begin{eqnarray}
t_{not}^{(n)}=(2n+1)\frac{\pi}{\omega} \qquad (n\in\mathbb{N})
\end{eqnarray}
for the following initial conditions
\begin{eqnarray}
(p_0,q_0)&=&(\forall,\mp 1), \\
\mathbf{S}_0&=&(0,0,\pm 1) \qquad (\textrm{poles}), \\
|\psi_0\rangle&=&|\pm\rangle,
\end{eqnarray}
and
\begin{eqnarray}
(p_0,q_0)&=&(l\pi,\forall) \qquad (l=0,1,2), \\
\mathbf{S}_0&=&\left((-1)^l\sqrt{1-q_0²},0,-q_0\right), \\
|\psi_0\rangle&=&\sqrt{\frac{1-q_0}{2}}|+\rangle +(-1)^l\sqrt{\frac{1+q_0}{2}}|-\rangle.
\end{eqnarray}
\end{subequations}

The set of initial conditions referent to the poles has already been reported in \cite{divicenzo95}. Notice that the points corresponding to the poles in the unit sphere are transformed in separatrices ($q=\pm 1$) in the phase space $qp$. It is just a manifestation in $pq$ of the ambiguity related to $\varphi$ in the poles of the unit sphere. 

The last set corresponds to the case of a {\em universal real} NOT-gate, since the initial quantum state is composed only by real coefficients. Note that there exist two different branches, depending on the value that $l$ assume. The branch $b0$, for $l=0,2\pi$, contains all initial condition satisfying $S_1>0$. On the other hand, for the branch $b\pi$ ($l=1$), one has $S_1<0$. These branches meet each other at $S_1=0$ ($q_0=\pm 1$, poles), onto which pass a separatrix. 

\vspace{0.3cm}
\noindent
{\bf CASE 3} \\ Provided
\begin{subequations}
\begin{eqnarray}
B_3&=&\frac{\omega}{2},\\
B_0&=&\omega,
\end{eqnarray}
the NOT operation occurs periodically for
\begin{eqnarray}
t_{not}^{(n)}&=&(2n+1)\frac{\pi}{2\omega} \qquad (n\in\mathbb{N}), \\
(p_0,q_0)&=&(\forall,\mp 1), \\
\mathbf{S}_0&=&(0,0,\pm 1)\qquad \textrm{(poles)}, \\
|\psi_0\rangle&=&|\pm\rangle,
\end{eqnarray}
and for
\begin{eqnarray}
t_{not}^{(n)}&=&(4n+\epsilon)\frac{\pi}{2\omega} \qquad (n\in\mathbb{N}),\\
(p_0,q_0)&=&\left([2l+3]\frac{\pi}{4},\forall\right), \\
\mathbf{S}_0&=&\left(\sqrt{1-q_0²}\cos p_0,\sqrt{1-q_0²}\sin p_0,-q_0\right),\qquad\,\,\, \\
|\psi_0\rangle&=&\sqrt{\frac{1-q_0}{2}} |+\rangle+\sqrt{\frac{1+q_0}{2}}e^{\imath p_0}|-\rangle,
\end{eqnarray}
in which
\begin{eqnarray}
\epsilon=\left\{\begin{array}{l} 1,\,\,\textrm{for $l$ even}, \\  3,\,\,\textrm{for $l$ odd}.  \end{array}   \right.
\end{eqnarray}
\end{subequations}
Note that in this case, $t_{not}$ depends on the choice for the $p_0$. Once again, there are several branches contributing to the NOT operation, namely, $b\frac{\pi}{4},\,b\frac{3\pi}{4},\,b\frac{5\pi}{4}$ and $b\frac{7\pi}{4}$, each one of them corresponding to a given $l$. These cases correspond to quantum states with complex phases in the form $e^{\imath p_0}=(\pm 1\pm\imath)/\sqrt{2}$, for all combinations of signals, for any $q_0$. Therefore, also in these cases, there exist a type of universality for the NOT operation.

\vspace{0.3cm}
\noindent
{\bf CASE 4} \\ With
\begin{subequations}
\begin{eqnarray}
B_3&=&\frac{\omega}{2},\\
B_0&=&\frac{\omega}{4 m} \qquad (m\in\mathbb{N^*}),
\end{eqnarray}
the NOT operation occurs periodically for
\begin{eqnarray}
t_{not}^{(n)}&=&m(2n+1)\frac{2\pi}{\omega} \qquad (n\in\mathbb{N}), \\
(p_0,q_0)&=&\left([l+\frac{1}{2}]\pi,\forall\right), \\
\mathbf{S}_0&=&(\sqrt{1-q_0²},0,-q_0), \\
|\psi_0\rangle&=&\sqrt{\frac{1-q_0}{2}}|+\rangle+\imath \sqrt{\frac{1+q_0}{2}}|-\rangle.
\end{eqnarray}
\end{subequations}
This case corresponds to the branches $b\frac{\pi}{2}$ and $b\frac{3\pi}{2}$.

\begin{figure}[ht]
\includegraphics[scale=0.3,angle=0]{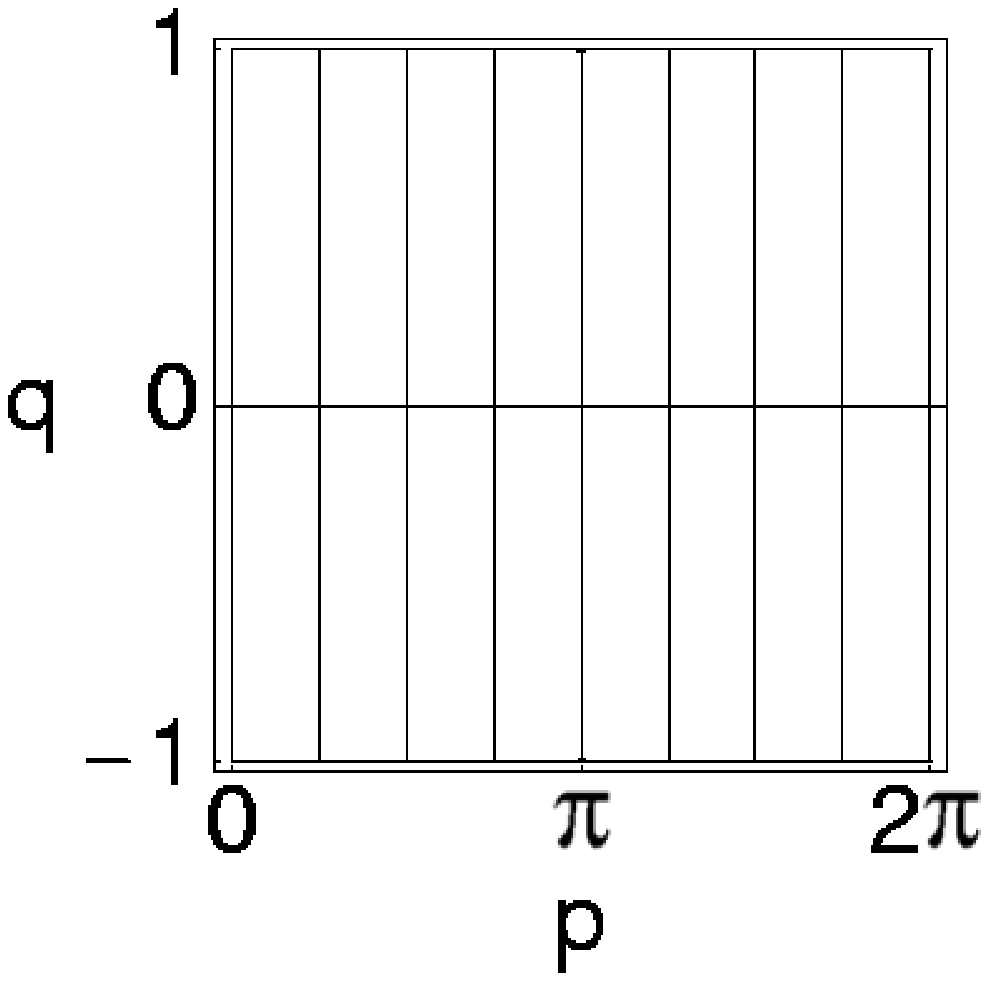}
\includegraphics[scale=0.6,angle=0]{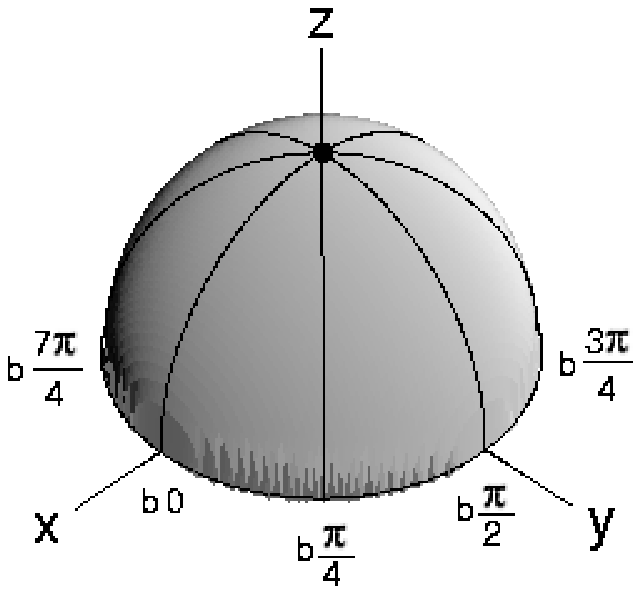}
\caption{\small Set of initial conditions able that allows for the implementation of the NOT operation in the R system (solid lines and the points corresponding to the poles of the sphere). The solutions are shown in both the phase space $pq$ (on the left) and the tridimensional phase space $xyz$. By simplicity, only the north hemisphere is shown, since the south presents the same contents.}
\label{solutions}
\end{figure}

For completeness, we point out that all informations got above concerning the initial conditions emerged from the following simple expressions derived from \eqref{SSrwa} at a specific $t_{not}$:
\begin{subequations}
\begin{eqnarray}
(1)\,\,\mathbf{S}_{not}\cdot\mathbf{S}_0&=&1-2 q_0²,\\
(2)\,\,\mathbf{S}_{not}\cdot\mathbf{S}_0&=&-q_0²+(q_0²-1)\cos\left(2 p_0 \right),\\
(3)\,\,\mathbf{S}_{not}\cdot\mathbf{S}_0&=&-q_0²+(-1)^{l+1}(q_0²-1)\sin\left(2 p_0 \right),\qquad\\
(4)\,\,\mathbf{S}_{not}\cdot\mathbf{S}_0&=&-q_0²+(1-q_0²)\cos\left(2 p_0 \right),
\end{eqnarray}
\end{subequations}
in which we have to impose that $\mathbf{S}_{not}\cdot\mathbf{S}_0=-1$. Figure \ref{solutions} presents an overview of the global symmetry presented by the set initial conditions allowed to implement the unitary NOT operation.

It is interesting to note that the initial conditions found in the case 4 are just localized over the separatrix of motion in phase space $pq$. However this is not relevant to the quantum dynamics, since the separatrix exists only in this particular phase space. It occurs in virtue of the topology change from a tridimensional phase space (sphere) to a bidimensional one (plane $qp$). A deeper analysis of the classical picture associated to the quantum dynamics may be found in appendix \ref{app:Gpicture}.

\subsection{NOT operation in the NR system}

\begin{figure}[ht]
\centerline{\includegraphics[scale=0.35,angle=-90]{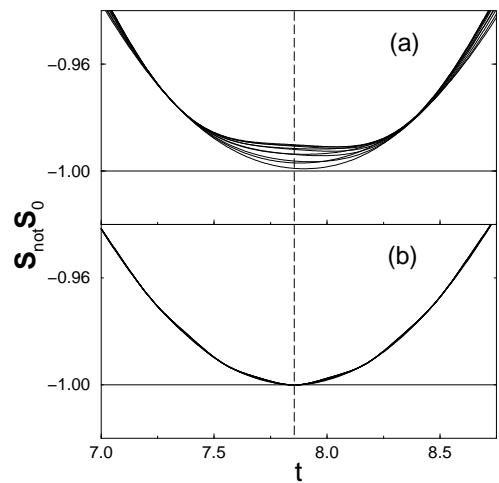}}
\caption{\small NR NOT operation with $B_0=B_3=0.2$ and (a) $\omega=3.0$ or (b) $\omega=10.0$. $\gamma$ tends to $2 B_3$ asymptotically, thus satisfying the resonance conditions of the case 2. However, here the universality class is $(p_0,q_0)=\left(\frac{3\pi}{2},\forall \right)$. The vertical dashed line stands for the analytical NOT instant given by $t_{not}=7.854$ (see appendix \ref{app:Gpicture}).}
\label{case2}
\end{figure}

The precedent analysis shows that practically all NOT regimes occurs for the resonance $B_3=\frac{\omega}{2}$, which produces stroboscopic maps with centered periodic orbits. The connection between R and NR maps verified in section \ref{sec:NR} indicates that we have to search regimes such that $B_3=\frac{\gamma}{2}$ in the NR system, since the parameter $\gamma$ plays the role of the R frequency $\omega$ for the map symmetry (see equations \ref{Smap} and \ref{NSmap}). According to \eqref{<f>} and \eqref{gammaf} this regime will be achieved only asymptotically, provided $\omega²(B_0²+B_3²)^{-1}\to 0$. We have verified numerically for the case 2 (Fig. \ref{case2}) that the NOT operation indeed occurs asymptotically, i.e., $\mathbf{S}_{not}\mathbf{S}_{0}\to-1$ as $\frac{B_0}{\omega}\to 0$, with $B_3=B_0$.

\begin{figure}[ht]
\includegraphics[scale=0.25,angle=-90]{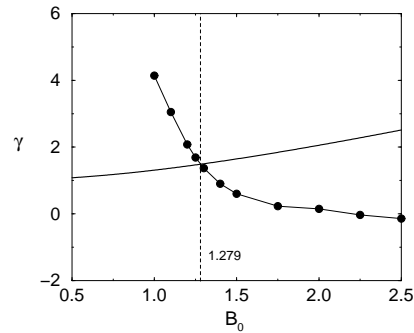}
\caption{\small Numerical determination of the $B_0$ value that satisfies \eqref{res} (tick solid line) for $\omega=1.0$ and $B_3=1.5$. For each value of $B_0$ we determine $\gamma$ (bullets) by means of the fitting procedure described in section \ref{sec:NR}.}
\label{gammaB0}
\end{figure}

\begin{figure}[ht]
\includegraphics[scale=0.3,angle=-90]{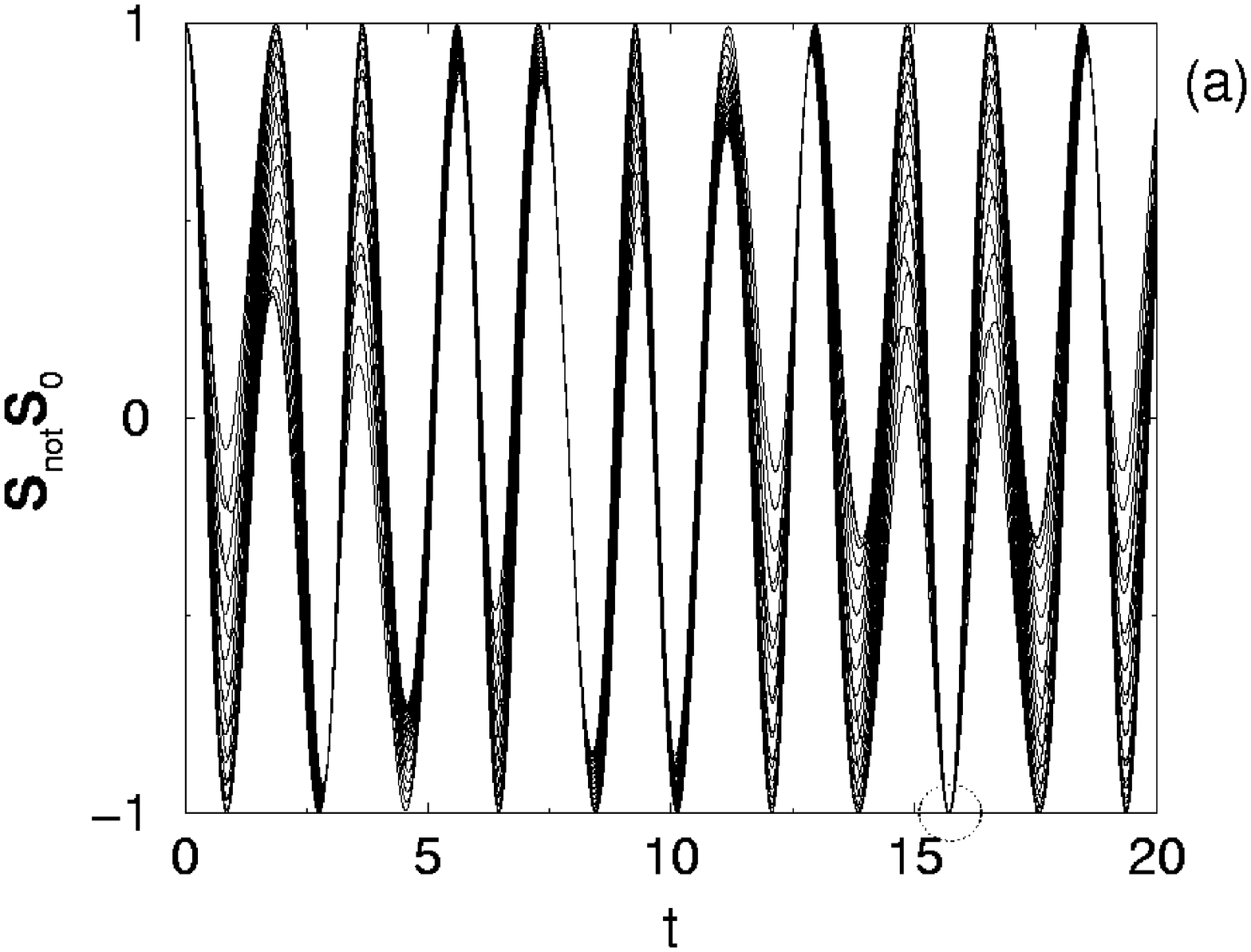}
\includegraphics[scale=0.3,angle=-90]{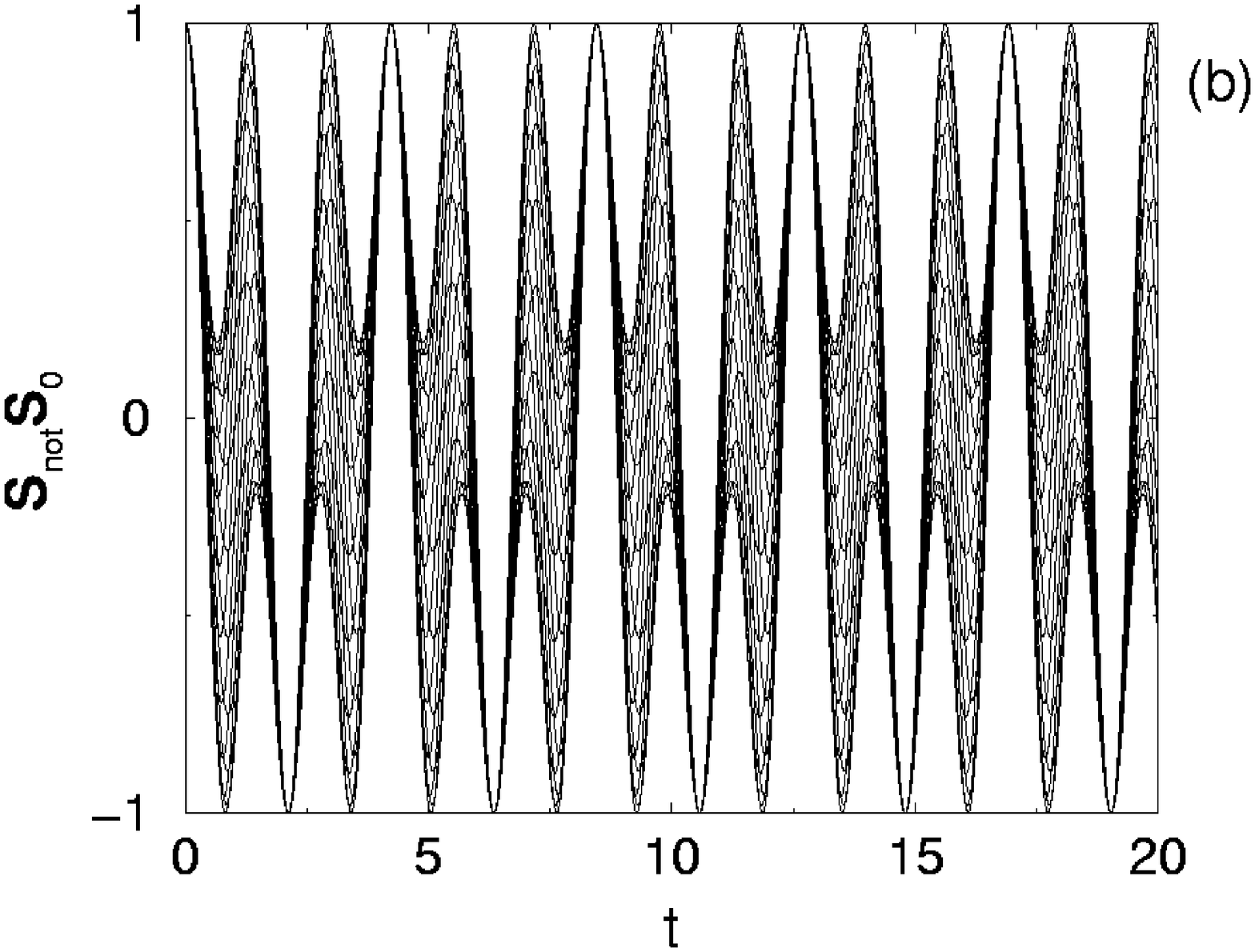}
\caption{\small (a) NR NOT operation for $\omega=1.0$, $\gamma=1.486$, $B_3=1.5$ and $B_0=1.279$ occurring at $t_{not}=5\pi$ for several initial conditions with $p_0=\frac{\pi}{2}$. (b) R NOT operation for $\omega=1.486$, $B_3=1.5$ and $B_0=1.279$, with $t_{not}=(2n+1)\frac{\pi}{\omega}$ for initial conditions given by $(p_0,q_0)=(\forall,0)$ (case 1).}
\label{ss}
\end{figure}

The most interesting and non-trivial regime of NR NOT operation was found at a resonance equivalent to the case 1. Obeying the analogy between $\gamma$ and $\omega$ we obtain the following resonance condition for the NR system:
\begin{eqnarray}
\gamma²=B_0²+\left(B_3-\frac{\gamma}{2} \right)²,\label{res}
\end{eqnarray}
being $\gamma$ a parameter determined by the choices of the NR frequencies $\omega$, $B_3$ and $B_0$, i.e., $\gamma=\gamma(\omega, B_0,B_3)$. We determined numerically the parameters satisfying \eqref{res} by fixing $\omega$ and $B_3$ and varying $B_0$. The result is shown in Fig. \ref{gammaB0}.
Once found the solutions for equation \eqref{res}, we verified the realization of the NOT operation for the NR system for a universality class given by $(p_0,q_0)=\left(\left[l+\frac{1}{2} \right]\pi,\forall \right)$, with $l\in\mathbb{N}$, as can be seen in Fig. \ref{ss}-(a). We also plot the R NOT operation with analog parameters.

Although not shown by the figure, the NR NOT operation is also periodic, with $t_{not}=(2n+1)5\pi$. Notice that the first occurrence of the NOT operation happens in the NR system five times latter than in the R system. This is an adverse point for the NR system, since decoherence effects may be not negligible in this time scale. 

At last we point out that the prediction of NOT operations in the NR system was based on the comparison of stroboscopic maps with those of the R system. These classical mathematical tools, based on the concept of trajectories flowing in the phase space, were crucial in the determination of the connection between the R frequency $\omega$ and its NR counterpart $\gamma$, parameters that define the symmetry of the maps. It stresses the usefulness of the classical gyromagnetic picture, being an example of classical analysis allowing quantum predictions.
\section{Summary and Final Remarks}\label{sec:conclusion}

We have presented some relevant results concerning the underlying classical aspects referent to the two-level systems under time-dependent external fields. The main results were obtained in virtue of the application of the classical gyromagnet picture, which has been shown to be a powerfull approach to investigate quantum two-level dynamics. 

We have shown that the underlying classical dynamics is integrable for arbitrary external fields as a consequence of the unitary quantum dynamics. The numerical stroboscopic maps, which are in perfect agreement with our analytical proof of integrability, were crucial mainly to the identification of the similarities between the R and NR dynamics. On the other hand, the average weighed up by the potential and the application of the averaging method allowed us to identify several interesting regimes with explicit analytical solutions.

It also should be remarked that we have arrived at other two results of independent interest: (i) it has been established beyond doubt (although not by rigorous proofs) that both the RWA and the averaging theorem are valid for such systems well outside their expected regions of validity; (ii) we have shown that the RWA is a formal consequence of the averaging theorem.

Concerning the realization of the NOT-gate in two-level systems, we have found several regimes for its experimental implementation for two types of external periodic fields. For a rotating field, the quantum dynamics has been solved and universality classes for the initial quantum state have been identified analytically, thus indicating all control parameters for the NOT-gate working. On the other hand, for the nonrotating field, analytical equations of motion were not found and predictions about the NOT operations were possible thanks to the classical analysis based on stroboscopic maps.

\acknowledgments

R.M.A. thanks financial support from FAPESP (Fundação de Amparo à Pesquisa do Estado de São Paulo) under grant 02/10442-6. W.F.W. thanks H.R. Jauslin and D. Sugny for remarks and suggestions.

\appendix
\section{Analytical solutions}\label{app:analytical}

As we mentioned, the solutions for the rotating field treated in section \ref{sec:R} were found by solving the quantum problem. Applying usual unitary transformations, namely $\exp{(-\imath\,\omega t \sigma_3)}$, we led the system to the rotating frame, in which the system is no longer time-dependent. Then the solution was promptly obtained. The last step consisted in applying the parametrization for classical canonical variables, $\mathbf{S}=(S_1,S_2,S_3)$ and $(p,q)$. The final solutions, including that one for $\mathbf{S}(t)\cdot\mathbf{S}(0)$, are given bellow as functions of the initial condition $(p_0,q_0)$, with $B=2\sqrt{B_0²+\Omega²}$ and $\Omega=B_3-\frac{\omega}{2}$.

\begin{widetext}
\begin{eqnarray}
S_1(t)&=&-\frac{4\,q_0\,B_0}{B²}\,\left[ 2\,\Omega \,\cos (w\,t)\,{\sin (B\,t/2)}^2 +  \frac{B}{2} \,\sin (w\,t)\,\sin (B\,t ) \right] + \frac{\sqrt{1 -q_0^2}}{B²}\,\Big[ 2\,B_0² \,\cos (p_0+w\,t) + \nonumber \\ && \left. + {\left( \frac{B}{2}  - \Omega  \right) }^2\,\cos (p_0+w\,t - B\,t ) + {\left( \frac{B}{2}  + \Omega  \right) }^2\,\cos (p_0+w\,t + B\,t ) + 4\,{B_0}^2 \,\cos (p_0-w\,t)\,{\sin (B\,t/2 )}^2 \right]\label{S1rwa} \\ \nonumber \\
S_2(t)&=&-\frac{4\,q_0\,B_0}{B²}\,\left[ 2\,\Omega \,\cos (w\,t)\,{\sin (B\,t/2 )}^2 - \frac{B}{2} \,\sin (w\,t)\,\sin (B\,t ) \right]+\frac{{\sqrt{1 - {{q_0}}^2}}}{B²}\,   \Big[ 2\,B_0² \,\sin (p_0+w\,t) + \nonumber \\ && \left. + {\left( \frac{B}{2}  - \Omega  \right) }^2\,\sin (p_0+w\,t - B\,t ) + {\left( \frac{B}{2}  + \Omega  \right) }^2\,\sin (p_0+w\,t + B\,t ) - 4\,{B_0}^2 \,\sin (p_0-w\,t)\,{\sin (B\,t/2 )}^2 \right] \label{S2rwa} \\ \nonumber \\
S_3(t)&=&-\frac{4\,q0}{B²}\left[ {\Omega }^2 + \cos (B\,t )\,{B_0}^2 \right] +\frac{4\,{B_0}\,{\sqrt{1 - {{q_0}}^2}}}{B²} \Big\{\,\Omega \,\cos ({p_0})\,\Big[1-\cos (B\,t)\Big] + \frac{B}{2} \,\sin ({p_0})\,\sin (B\,t )\Big\}. \label{S3rwa}
\end{eqnarray}
\begin{eqnarray}
q(t)&=&-S_3(t). \\ \label{qrwa}
p(t)&=&\arctan{\left[\frac{S_2(t)}{S_1(t)}\right]}+\left\{
\begin{array}{l}
0, \qquad \textrm{if}\,\,S_1(t)\ge 0; \\ \\
\pi, \qquad \textrm{if}\,\,S_1(t)< 0.
\end{array}  \right. \label{prwa}
\end{eqnarray}
\begin{eqnarray}
\mathbf{S}(t)\cdot\mathbf{S}(0)&=&\frac{2 \,B_0²\,\cos (w\,t) + 2\,
    \left( \frac{B²}{4} + {\Omega }^2 \right)\cos (\omega\,t)\,\cos (B\,t) \,- 2\,B \,\Omega \,\sin (w\,t)\,\sin (B\,t )
    }{B²} +\nonumber \\ && - \frac{ 
    8\,q_0²\,\Big[ \frac{B}{2} \,\cos (B\,t/2 )\,\sin (\frac{w\,t}{2}) + 
         \Omega \,\cos (\frac{w\,t}{2})\,\sin (B\,t/2 ) \Big]^2 }{B²} + \nonumber \\ && -\frac{ 
    8\,{B_0}\,{q_0}\,
     \sqrt{1 - q_0^2}\cos (p_0-\frac{w\,t}{2})\,\left[ 2\,\Omega \,\cos (\frac{w\,t}{2})\,{\sin (B\,t/2)}^2 + 
       \frac{B}{2}\,\sin (\frac{w\,t}{2})\,\sin (B\,t ) \right] }{B²} +\nonumber \\ &&+\frac{4\,{B_0}^2 \,{\sin (B\,t/2 )}^2\,
     \left\{ \cos (2\,p_0-w\,t) -{q_0}^2\, \Big[ 1 + \cos (2\,p_0-w\,t) \Big] \right\}}{B²}. \label{SSrwa}
\end{eqnarray}
\end{widetext}

\section{Geometric Picture} \label{app:Gpicture}

The classical version of to the two-level quantum system may be understood as a {\em classical gyromagnet}, with $\mathbf{S}$ precessing around the external field $\mathbf{B}$. In this appendix this analogy is emphasized by means of a geometrical picture of the classical dynamics, which illustrates questions concerning the separatrix, the periodic orbits and the NOT operation.

We start our analysis by focusing on the R system in its rotated version, namely, 
\begin{subequations}
\begin{eqnarray}
\cal{K}&=&-\mathbf{B}\cdot\mathbf{S}=-B \,\cos{\Psi}, \label{KPsi} \\
\mathbf{B}&=&(B_x,B_y,B_z)=-2\,(B_0,0,\Omega),\label{Bconstant} \\
\mathbf{S}&=&(\sqrt{1-Q²}\cos P,\sqrt{1-Q²}\sin P,-Q),
\end{eqnarray}
\end{subequations}
with $B=||\mathbf{B}||$ and being $\Psi$ the angle between the external field and the unit vector $\mathbf{S}$. The dynamics is governed by
\begin{eqnarray}
\mathbf{v}\equiv\frac{d\mathbf{S}}{dt}=\mathbf{S}\times\mathbf{B},
\label{v}
\end{eqnarray}
which defines the velocity of the unit vector $\mathbf{S}$.
We may calculate the acceleration $\mathbf{a}=\mathbf{\dot{v}}$ noting that the external field is time-independent. This yields
\begin{eqnarray}
\mathbf{a}=\mathbf{v}\times\mathbf{B}.
\label{a}
\end{eqnarray}

Consider two orthogonal versors, say $\mathbf{e}_{\|}$ (parallel to the $\mathbf{B}$) and $\mathbf{e}_{\bot}$ (perpendicular to $\mathbf{B}$), such that we may decompose: $\mathbf{S}=\cos{\Psi}\,\,\mathbf{e}_{\|}+\sin \Psi \,\mathbf{e}_{\bot} $ and $\mathbf{B}=B\,\mathbf{e}_{\|}$. Since $\cal{K}$ and $\mathbf{B}$ are constants the only option for $\mathbf{S}$ is performing a precession around the static magnetic field. This consideration makes the versor $\mathbf{e}_{\bot}$ time-dependent.

Consequently, the velocity and acceleration can be written respectively as
\begin{subequations}
\begin{eqnarray}
\mathbf{v}&=&B\,\sin \Psi \,\,\left(\mathbf{e}_{\bot}\times\mathbf{e}_{\|} \right),\label{vPsi}\\
\mathbf{a}&=&B²\,\sin \Psi\,\,\left(-\mathbf{e}_{\bot}\right).\label{aPsi}
\end{eqnarray} \label{avPsi}
\end{subequations}
Being $\cal{K}$ and $\mathbf{B}$ constants, we conclude by \eqref{KPsi} that $\Psi$, $\textrm{v}=||\mathbf{v}||$ and $\textrm{a}=||\mathbf{a}||$ are all constants too. Moreover, \eqref{aPsi} indicates that $\mathbf{a}$ is centripetal, stressing the fact that $\mathbf{S}$ develops a precession motion around $\mathbf{B}$.

By \eqref{KPsi} we note that $\Psi \in [0,\pi]$ and $\cal{K} \in [-B,B]$. This allows us to re-write equations \eqref{avPsi} in terms of the energy as
\begin{subequations}
\begin{eqnarray}
\mathbf{v}&=&\sqrt{B²-\cal{K}²}\,\,\left(\mathbf{e}_{\bot}\times\mathbf{e}_{\|} \right),\\
\mathbf{a}&=&B \sqrt{B²-\cal{K}²}\,\,\left(-\mathbf{e}_{\bot}\right).
\end{eqnarray}
\end{subequations}

Finally, we may define the angular velocity $\mathbf{w}=(\sin\Psi\,\mathbf{e}_{\bot})\times \mathbf{v}$ which defines the precession direction. Using equations given above we get
\begin{eqnarray}
\mathbf{w}&=&B\sin²\Psi \left(-\mathbf{e}_{\|}\right)=-\left(B²-\cal{K}² \right)\mathbf{e}_{\|},
\end{eqnarray}
which shows that the angular velocity is always anti-parallel to $\mathbf{B}$. This helps us to understand the direction of the trajectories flux. Moreover, since $\textrm{w}=||\mathbf{w}||$ is constant we may define a rotation angle as $\Phi(t)=\Phi_0+\textrm{w} t$.

\subsection{Periodic Orbits (Eigenstates)}

A first interesting consequence of the simple scheme developed up to here is that the initial condition corresponding to the periodic orbits (eigenstates) of the system may be promptly obtained for  $\cal{K}²=B²$, thus yielding $\mathbf{v}=\mathbf{a}=0$. Therefore, these periodic orbits correspond to ``static'' quantum states (eigenstates). In this situation, $\mathbf{S}$ and $\mathbf{B}$ are parallel $(\Psi=0,\pi)$, and there is no external torque.

\subsection{Separatrix} \label{sub:separatrix}

\begin{figure}[ht]
\includegraphics[scale=0.275,angle=0]{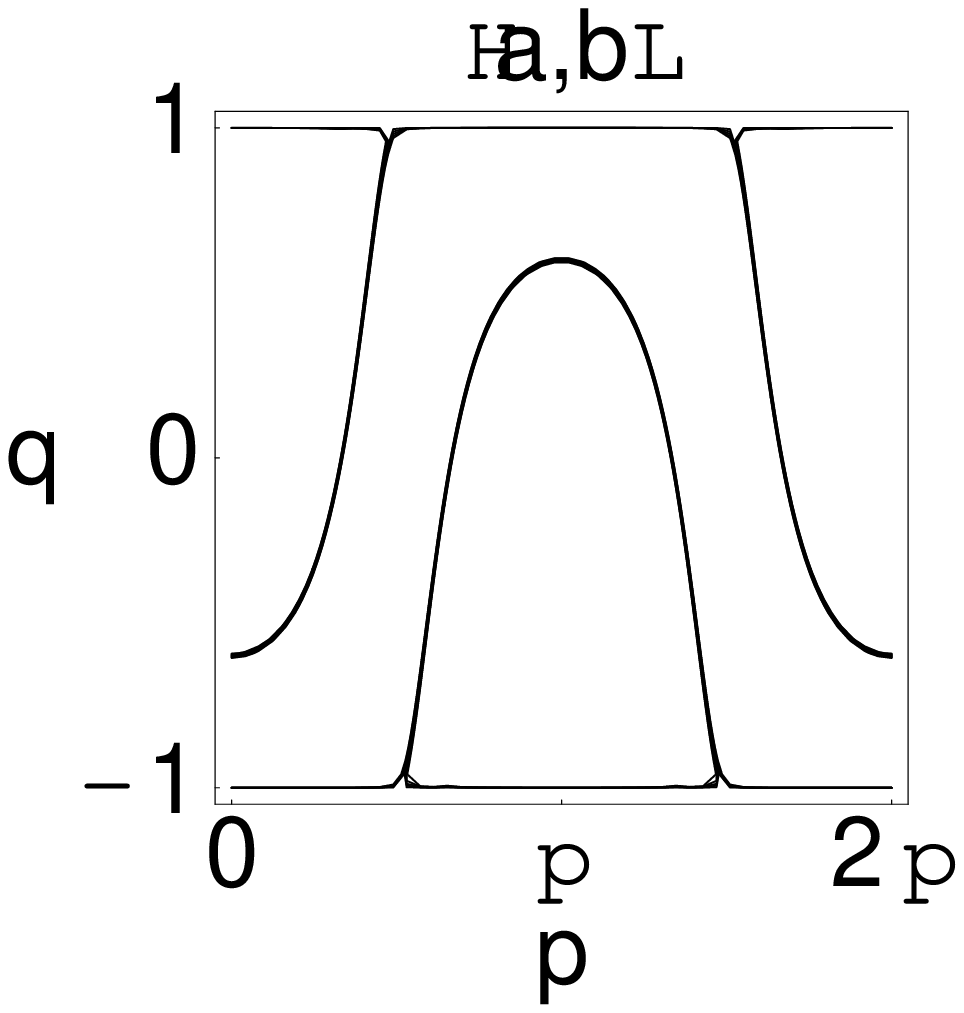}
\includegraphics[scale=0.275,angle=0]{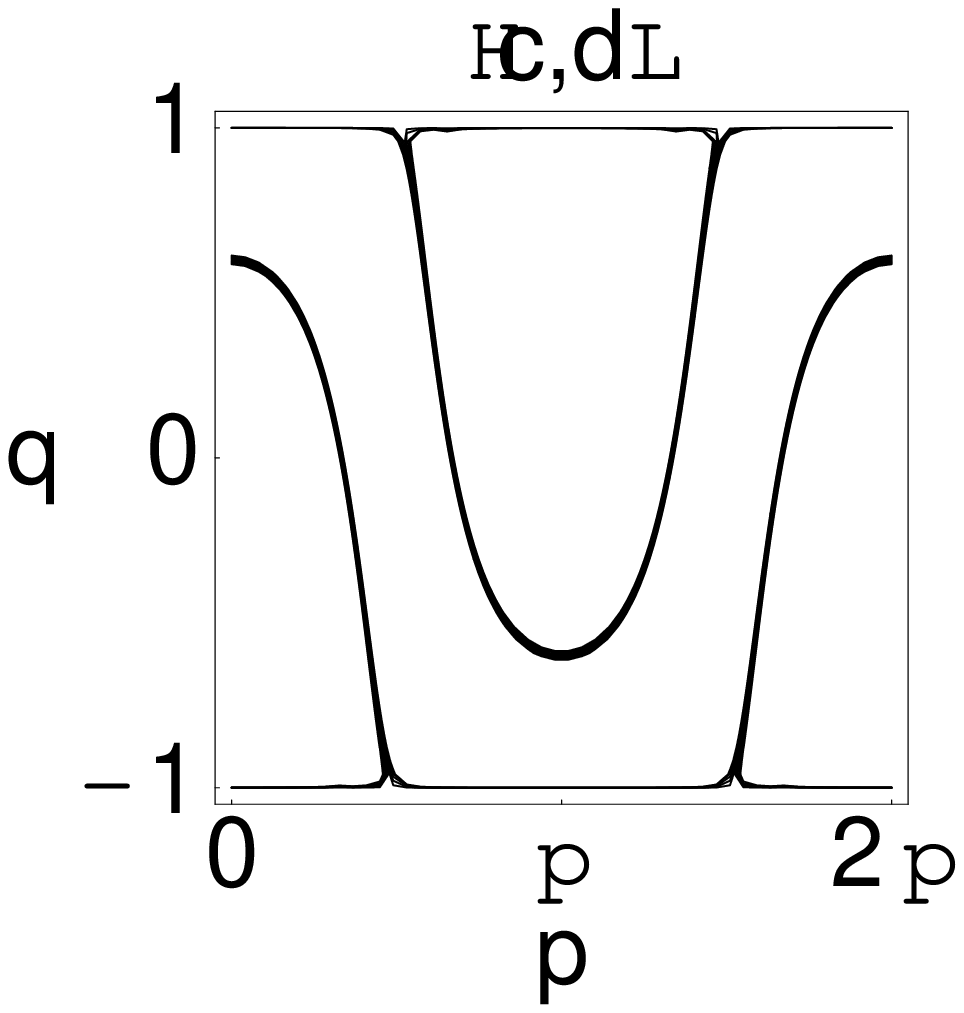}
\includegraphics[scale=0.275,angle=0]{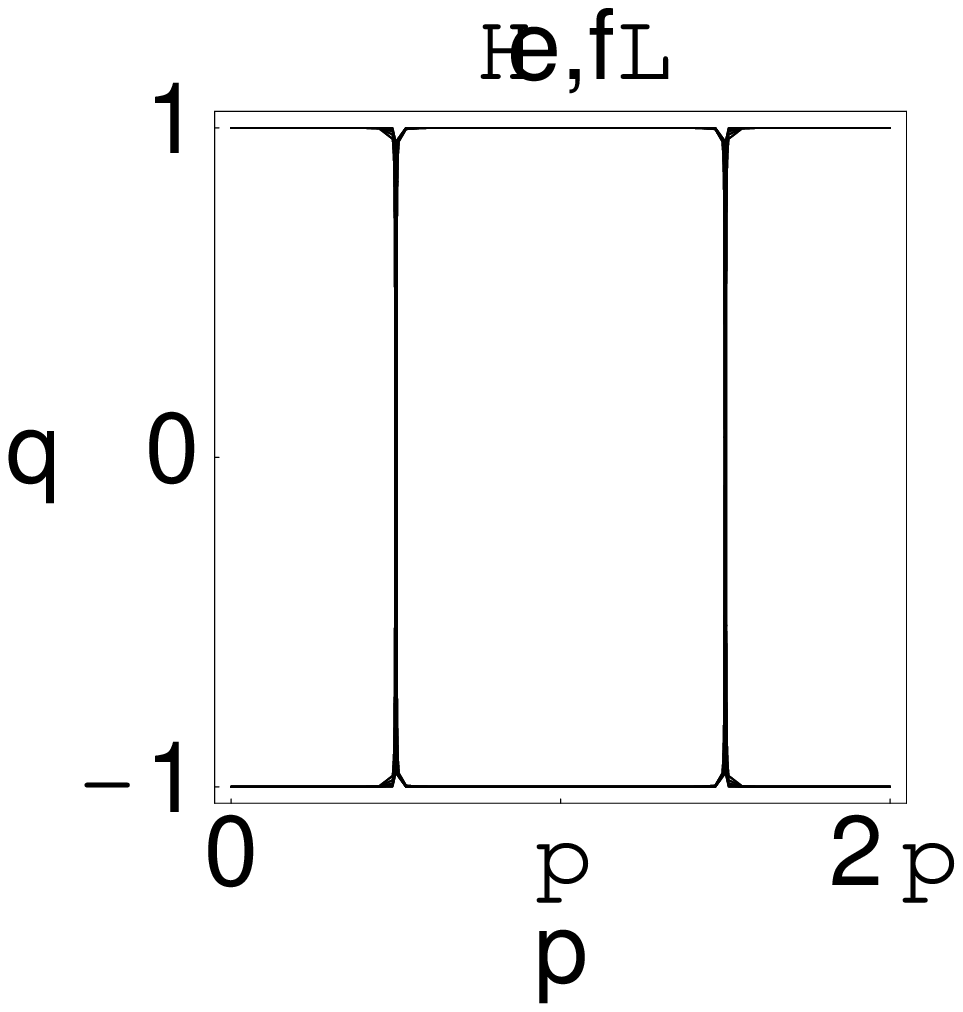}
\caption{\small Separatrices of motion plotted with the same parameters of the respective graphics shown in Fig. \ref{fig1}.}
\label{fig:separatrix}
\end{figure}

The classical gyromagnet presents in the space $pq$ a separatrix of motion associated to the specific energy $\pm 2\Omega$. Therefore, by \eqref{Smap} with $\Omega=B_3-\frac{\omega}{2}$ and $\phi=0$, the explicit contour for the separatrix at stroboscopic instant is given by
\begin{eqnarray}
B_0 \sqrt{1-q_k²}\cos p_k=(\pm 1+q_k)\, \Omega.
\label{separatrix}
\end{eqnarray}
This equation offers all forms for the separatrix, as is shown in Fig. \ref{fig:separatrix}.

We then have $\cal{K}=\pm B_z$, in which the signals indicate the direction of the vector $\mathbf{S}$ compared to $\mathbf{B}$ fixed. Consider now that $\Theta$ is the projection angle of the magnetic field along the direction $z$, i.e., $B_z=B\cos{\Theta}$. Hence, according to \eqref{KPsi} we obtain $\Psi=\Theta$ or $\Psi=\pi-\Theta$, which indicates that the vector $\mathbf{S}$ will be necessarily parallel (or anti-parallel) to the versor $\mathbf{z}$ at some instant. Therefore, inevitably $\mathbf{S}$ will pass through the poles during its precession motion. However, concerning the dynamics of the unit vector $\mathbf{S}$ on the sphere, there is absolutely nothing special in this situation. For instance, the period of the precession in this case is
\begin{eqnarray}
\tau=\frac{2\pi}{\textrm{w}}=\frac{2\pi}{B_x²}=\frac{\pi}{2 B_0²},
\end{eqnarray}
which is remarkably smaller than the period (infinity) expected for a separatrix. In this sense, there does not exist a separatrix in the classical unit sphere. It appears only in the $pq$ (or $PQ$) phase space reflecting the peculiarities of its topology, which transforms ``localized'' states on the sphere (poles) into ``delocalized'' ones on the plane (lines $(p,q)=(\forall,\pm 1)$).

\begin{figure}[ht]
\includegraphics[scale=0.5,angle=0]{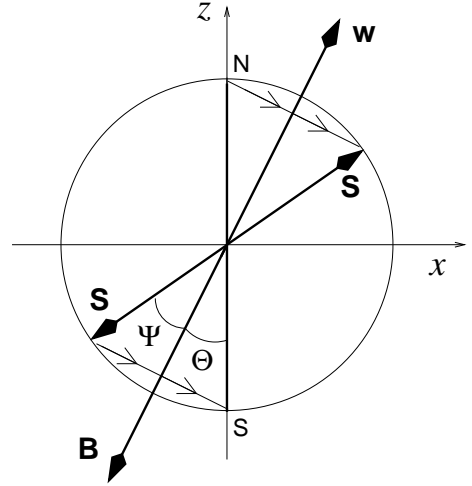}
\caption{\small Unit sphere projected in the plane $xz$. For the separatrix energy $\cal{K}=\pm B_z$ the vector $\mathbf{S}$ precesses around $\mathbf{B}$ with $\Psi=\Theta$. In this situation the unit vector crosses the poles of the sphere, which corresponds to the unstable points of the space $PQ$. The angular velocity $\mathbf{w}$ and the flux direction are also indicated.}
\label{precession}
\end{figure}

In addition, we note that the hyperbolic fixed points $(\frac{\pi}{2},\pm 1)$ by the which the separatrix passes through, have no special physical interpretation on the sphere. Figure \ref{precession} illustrates geometrically our argumentation.

\subsection{NOT operation}

Let us work with matrix notation, using $S_R^T=(S_x\,\,S_y\,\,S_z)$ and $S_r^T=(S_1\,\,S_2\,\,S_3)$ for the line matrices corresponding to the unit vectors in phase spaces $PQ$ (index $R$) and $pq$ (index $r$) respectively. The superscript $T$ denotes the transposition operation. According to the canonical transformation \eqref{F3} we may write
\begin{subequations}
\begin{eqnarray}
S_R(t)=G(t)S_r(t),
\end{eqnarray}
\begin{eqnarray}
G(t)=\left(
\begin{array}{ccc}
\cos{\omega t} & \sin{\omega t} & 0 \\
-\sin{\omega t} & \cos{\omega t} & 0 \\
0 & 0 & 1 
\end{array} \right),
\end{eqnarray}
\end{subequations}
being $G(t)$ a rotation matrix around the axes $z$ (or ``3''). Consequently, $G^{-1}(t)=G^{T}(t)=G(-t)$ and $S_r(t)=G(-t)S_R(t)$. Now, it is easy to see that
\begin{subequations}
\begin{eqnarray}
S_R^T(t)\,S_R(0)&=& S_r^T(t)\,G(-t)\,S_r(0),\\
S_r^T(t)\,S_r(0)&=& S_R^T(t)\,G(t)\,S_R(0).
\end{eqnarray}
\end{subequations}
This result shows that in general the NOT operation does not occur simultaneously in both frame.

The NOT operation instant $\tau_{not}$ in the frame $R$ may me determined from the geometrical picture developed in this appendix. It reads $\tau_{not}=\frac{\pi}{\textrm{w}}=\frac{\pi}{B}$, which is a half of the precession period with $\Psi=\frac{\pi}{2}$. This value for $\Psi$ is a necessary condition for the NOT operation in the frame $R$.  

All regimes found for the NOT operation in section \ref{NOTgate} may be formulated in terms of a single relation. To get it we re-write Eq.\eqref{KPsi} at $t=0$ as
\begin{eqnarray}
B\cos{\Psi}= B_x S_{1}(0)+ B_z S_3(0),
\end{eqnarray}
in which $S_1(0)=S_x(0)$ and $S_3(0)=S_z(0)$, since $(P_0,Q_0)=(p_0,q_0)$. In all regimes identified we have either $S_3(0)=0$ ($q_0=0$) or $B_z=0$ (resonance $B_3=\omega/2$). Then, using the projection angle $\Theta$ defined in section \ref{sub:separatrix}, we get
\begin{eqnarray}
\cos{\Psi}= S_1(0)\,\sin{\Theta},
\label{NOTrule}
\end{eqnarray}
with $S_1(0)=\sqrt{1-q_0²}\cos{p_0}$. Relation \eqref{NOTrule} puts together informations about the precession angle $\Psi$ (related to the energy), the magnitudes of the magnetic field projections (related to resonances) and the initial conditions (related to the initial quantum state). All regimes of the NOT operation obey this simple formula.

By \eqref{NOTrule} we see that the NOT operation will occur in both frames either if $(p_0,q_0)=\left(\frac{\pi}{2},\forall\right)$ or if $(p_0,q_0)=(\forall,\pm 1)$, since $B_0$ and $\Theta$ are always different of zero. This conditions are verified in cases 2, 3 and 4. Furthermore, in theses cases $B_3=\omega/2$, so $B=2 B_0$ and $\tau_{not}=\frac{\pi}{2 B_0}=t_{not}$. Therefore, the NOT operation will occur {\em simultaneously} in both frames.

\subsection{Mean NR NOT operation}

Let us consider the NR system, for the which the magnetic field is given by $\mathbf{B}(t)=-2(B_0,0,B_3\cos{\omega t})$. In the high frequency limit we may apply the averaging theorem to get
\begin{subequations}
\begin{eqnarray}
\bar{\cal{H}}=\bar{\mathbf{B}}\cdot\mathbf{S}
\end{eqnarray}
\begin{eqnarray}
\bar{\mathbf{B}}=\frac{1}{T}\int\limits_{0}^{T} \mathbf{B}(t)\, dt=(-2B_0,0,0).
\end{eqnarray} \label{meanH}
\end{subequations}
We see that in average, the field behaves like a time-independent external perturbation, allowing us to use the gyromagnet picture developed above. Thus, we may describe the precession motion by means of a constant average angle $\bar{\Psi}$ such that $\bar{\cal{H}}=-\bar{B}\cos{\bar{\Psi}}$, with $\bar{B}=||\mathbf{\bar{B}}||$. This equation may be written at the initial instant with \eqref{meanH} as
\begin{eqnarray}
\cos{\bar{\Psi}}=-\sqrt{1-q_0²}\,\cos{p_0}.\label{meanPsi}
\end{eqnarray}
Following the steps of the precedent sections one may determine the angular velocity. The result is simply $\textrm{w}=\bar{B}\sin{\bar{\Psi}}$. But we know that the NOT operation occurs only with $\bar{\Psi}=\frac{\pi}{2}$. In this situation, the NOT instant is directly calculated from the expression for $\textrm{w}$ as the time spent for performing a half rotation: $t_{not}=\frac{\pi}{\textrm{w}}=\frac{\pi}{\bar{B}}$. Moreover, in this regime equation \eqref{meanPsi} reads $\sqrt{1-q_0²}\,\cos{p_0}=0$, which is satisfied in general for $(p_0,q_0)=\left(\left[l+\frac{1}{2} \right]\pi,\forall \right)$. This result explain the universality class found in section \ref{NOTgate} for the NR system in the high frequency limit. Furthermore, using the numerical parameters of the Fig. \ref{case2} we get $t_{not}= 7.854$, in perfect agreement with the numerical results.


\end{document}